\documentclass[entropy,article,accept,manyauthors]{Definitions/mdpi}

\firstpage{1} 
\pubvolume{28}
\issuenum{1}
\articlenumber{123}
\pubyear{2026}
\copyrightyear{2026}
\externaleditor{Adam Lipowski} 
\datereceived{6 November 2025} 
\daterevised{12 January 2026} 
\dateaccepted{14 January 2026} 
\datepublished{20 January 2026 } 
\hreflink{https://doi.org/10.3390/e28010123} 

\Title{What Is a Pattern in Statistical Mechanics? 
Formalizing Structure and Patterns in One-Dimensional Spin Lattice Models with Computational Mechanics}

\Author{Omar 
Aguilar \orcidA{}}

\AuthorNames{Omar Aguilar}

\address[1]{%
Physics Department, University of California, Santa Cruz, 
 1156 High Street, Santa
Cruz, CA 95064, USA; omalagui@ucsc.edu}

\abstract{This work formalizes the notions of structure and pattern for three distinct one-dimensional spin-lattice models (finite-range Ising, solid-on-solid, and three-body), using information- and computation-theoretic methods. We begin by presenting a novel derivation of the Boltzmann distribution for finite one-dimensional spin configurations embedded in infinite ones. We next recast this distribution as a stochastic process, thereby enabling us to analyze each spin-lattice model within the theory of computational mechanics. In this framework, the process's structure is quantified by excess entropy
~$\textbf{E}$ (predictable information) and statistical complexity $C_{\mu}$ (stored information), and the process's structure-generating mechanism is specified by its $\epsilon$-machine. To assess compatibility with statistical mechanics, we compare the configurations jointly determined by the information measures and $\epsilon$-machines to typical configurations drawn from the Boltzmann distribution, and we find agreement. We also include a self-contained primer on computational mechanics and provide code implementing the information measures and spin-model distributions.}


\begin{document}


\section{\label{sec:intro}Introduction}

When observing a natural system, we intuitively explain it by describing the way its components are arranged. We might say that the system displays order or randomness. We might describe systems that exhibit a blending of order and randomness as complex or \textit{structured
} ({This paper uses two notions of structure. One refers to a system’s general type of arrangement, which we call \textit{generic} structure. The~other captures a more specific type of arrangement—one that exhibits patterns—which we call \textit{intrinsic} structure. Throughout the paper, the~intended notion will be clear from context.
}) \cite{aaronson2014coffee}. Moreover, we might also regard as structured those ordered systems that have no randomness but exhibit a repetition of more than one component (a period greater than 1) \cite{feldman1998dnco}. Altogether, we might regard a structured system simply as one that exhibits \textit{patterns} \cite{bak2013how}.

In light of this depiction, a~physicist may feel compelled to bring clarity and definiteness to the notions of randomness, structure and pattern by formalizing them. Although~statistical mechanics readily concretizes randomness through measures like entropy~\cite{rothstein1951, eliazar2021}, it falls short when quantifying structure and pattern and formalizing its supporting mechanism. For~instance, while magnetization is commonly treated as an indicator of structure, materials with distinct magnetic behaviors, such as paramagnets and antiferromagnets, have the same magnetization in the absence of a magnetic field: zero~\cite{feldman1998dnco}.

Furthermore, in~statistical mechanics, ``pattern'' is not typically defined explicitly; instead, the~criteria that one might regard as proxies for pattern depend on a choice of representation. This choice can enter through the observable taken as relevant, the~scale at which structure is probed, or~the coarse-graining scheme used to obtain a macroscopic description~\cite{sethna2006statmech}. For~the observable, one may diagnose order using magnetization or staggered magnetization~\cite{goldenfeld1992lectures}. For~the scale, one may use correlations or structure factors evaluated at a chosen length or wavenumber~\cite{chaikinlubensky1995pcmp}. For~coarse-graining, one may formalize large-scale organization through a specific RG blocking or decimation prescription~\cite{kadanoff1966scaling}.

Faced with these limitations, the~physicist may make their endeavor more concrete by posing two key questions:

\begin{enumerate}
    \item What's a simple system in statistical mechanics that manifests structure and patterns?
    \item How could one extend statistical mechanics to formalize structure and patterns within such a system?
\end{enumerate}

One-dimensional (1D) spin lattice models~\cite{yeomans1992} (p. 67) are suitable candidates for addressing these challenges, as~they compactly represent interacting magnets as spins in an evenly spaced grid, embodying both simplicity~\cite{krinsky1975} and structure/patterns~\cite{gheissari2019}. The~simplicity stems from the spins taking discrete values (often binary) and the spin models being amenable to both analytical and numerical treatment~\cite{schulz1999, lacombe1974}. The~structure and patterns are evident in the model's possible spin configurations, which exhibit regularity, randomness, and structure. For~example, the~1D nearest-neighbor Ising model may have configurations rich in regularity, randomness, and structure, such as $\uparrow \downarrow \uparrow \downarrow \uparrow \downarrow$, $\uparrow \downarrow \uparrow \downarrow \downarrow \downarrow \uparrow$ and $\downarrow \uparrow \uparrow \downarrow \uparrow \uparrow$, respectively. These configurations contain repeating sequences of spins that we refer to as \textit{configuration patterns}.

Mathematically, a~spin model is expressed as a Hamiltonian that characterizes the energy of the spin system~\cite{yeomans1992} (p. 67). Given the Hamiltonian, the~usual goal is to determine the partition function and from it compute various properties of interest~\cite{binder2015}. Among~these, the~Boltzmann distribution as a function of spin configurations is the least frequently computed ({When the Boltzmann distribution is calculated, 
 it is typically expressed as a function of energy~\cite{mccoy1968, beale1996} or other macroscopic properties~\cite{kofinger2010, tsypin2000, chatelain2006}, rather than directly in terms of configurations of fixed length}), yet it stands out as the sole one directly addressing spin configurations, serving as a window for analyzing their structure and patterns. However, to~clearly see through this window, we need to carefully consider how the distribution is~formalized.

Typically, the~Boltzmann distribution is defined so that each configuration, either implicitly or explicitly, represents an event of a single random variable, as~indicated in Refs.~\cite{pathria2011stat_mech} (p. 552) and~\cite{derrida1980}. Nonetheless, this approach is not conducive to examining how individual spins make up spin configurations. Instead, we can regard them as realizations of a partially ordered chain of random variables—a stochastic process~\cite{feldman1998dnco}.

In this process, which we call the spin process, each spin corresponds to an event of a single random variable. Given this perspective, we can now quantify the randomness, regularity, and~structure of the spin process, and~formalize the mechanism that generates its structure. Since randomness, regularity, and~structure are ways in which a process elicits surprise, we quantify them as information—a measure of ``quantifiable surprise''~\cite{tribus_thermostats_thermo} (p.~64) or a ``difference that makes a difference'' \cite{bateson1987steps}.

In information theory, the~theory of quantifiable surprise, a~stochastic process's intrinsic randomness or average randomness per symbol is quantified by its Shannon entropy rate $h_{\mu}$ (Ref.~\cite{cover2006info}, pp. 74--76). The~process’s regularity, as~the counterpart of its randomness, can be understood as the total correlation within the process. Thus, regularity is quantified as the amount of information that is shared within the process—that is, the~process's mutual information or excess entropy \textbf{E} \cite{shaw1984, crutchfield1983, grassberger1986, lindgren1988}. 

Because a stochastic process’s structure is effectively captured by its patterns, we quantify the process’s structure by measuring the amount of information stored in those patterns. This quantity is known as the stored information, or~statistical complexity $C_{\mu}$~\mbox{\cite{crutchfield1989inferring, crutchfield1994calculi}}, and~is defined as the Shannon entropy of those patterns. Calculating $C_{\mu}$, therefore, requires identifying these patterns—an inference task that effectively uncovers the process’s underlying structure-generating mechanism. We define these patterns~next.

Since patterns are sought for their predictive utility, we define a pattern in the spin process setup from a prediction‐based viewpoint. To~do so, we split ({Without loss of generality}) each realization of the spin process into a left half (past) and a right half (future). Then, we define a pattern as the set of pasts that \textit{lead} to the same futures ({It should be highlighted that for 1D spin lattice models, the~conventional time index is taken to be site location index and there is no time dependence.}) \cite{crutchfield2012btworderchaos}. By~``lead to”, we mean that the conditional distribution over futures, when conditioned on any past in the set, is identical across all those pasts. This condition is known as the causal equivalence principle ({This principle formalizes the implicit definition of a state commonly used in theoretical computer science when constructing machines. In~this context, a~state represents the information that must be retained to predict the system's future behavior (see Appendix~\ref{appendix:TOC_concept_of_state}).}) \cite{crutchfield1994calculi, shalizi2001compmech, crutchfield2012btworderchaos}, which recasts these patterns as causal states. Why the term ``state''? Because this conception of pattern is consistent with the theory of computation's definition of state as a system's entity that ``remembers a relevant portion of the system's history''~\cite{hopcroft2001} (pp. 2--3). This connection points us toward the mechanism that underpins the process’s~structure.

Given that a system's structure is measured in units of information, formalizing its supporting mechanism is tantamount to unraveling how the system processes and stores information—essentially, how it computes~\cite{shalizi1999thermo}. This leads to a refined question: what is the minimal~({To avoid accounting for computation not inherent to our system}) \textit{abstract machine}~({In the 21st century, ``computation” often evokes laptops, which perform \textit{useful computation}—that is, computation carried out for some external task. In~contrast, we focus on \textit{intrinsic computation}, the~computation a system performs by itself. To~analyze this, we use \textit{abstract machines} \cite{hopcroft2001}—mathematical models that consist of states and transitions and laid the groundwork for the theory of computation,}) which performs the computation inherent to the spin process. Leveraging concepts from the theory of computation (TOC), computational mechanics provides a compelling response: the set of causal states and their transitions, that is a $\epsilon$-machine or Probabilistic Deterministic Finite State Machine (PDFM). Here, ``probabilistic'' means that state transitions include probabilities, while ``deterministic'' implies that when we have knowledge of a state and its associated outgoing symbol, we have complete certainty about the next state we will transition to. Several methods have been developed for inferring $\epsilon$-machines~\cite{still2010optimal, streloff2014bayes, marzen2016predictive, rupe2019disco, rupe2020spacetime, brodu2022discovering, jurgens2025inferring}. Among~these, Feldman and Crutchfield's approach stands out as the only one that is both analytical and applicable to statistical mechanics~\cite{feldman1998dnco}.

In particular, Feldman and Crutchfield used this method to examine the structure of the nearest-neighbor and next-nearest neighbor Ising models. Subsequent research further developed their information-theoretic analysis of spin systems by calculating $h_{\mu}$ and $\textbf{E}$ for the two-dimensional nearest neighbor Ising model~\cite{feldman2003twodim}, as~well as decomposing the nn Ising model's Shannon entropy rate into more refined information components~\cite{vijayaraghavan2016anatomy}. Moreover, quantum $\epsilon$-machine formulations revealed striking memory advantages—ranging from extreme compression when simulating long-range Ising spin chains~\cite{aghamohammadi2017extreme} to clarifying how simplicity differs in quantum versus classical descriptions~\cite{aghamohammadi2017ambiguity}. Now, the~aim of this paper is to develop information measures and $\epsilon$-machines for three varied one-dimensional spin-lattice models—finite-range Ising, solid-on-solid, and~three-body—and to assess the consistency of these results with statistical~mechanics.

These developments are timely because they broaden the rapidly evolving landscape of abstract machines used to analyze computation in physical processes in two key ways. First, they encourage the application of abstract machines—which have most often been used to study thermodynamic~\cite{chattopadhyay2024thermomachine, chu2018thermodynamic, strasberg2015thermo, wolpert2024jordan} and quantum~\cite{li2024qmmachine, bhatia2019qmmachinesurvey, wang2019local, molina2019turingqmcircuit} processes—to statistical mechanical processes, potentially supporting more efficient information processing in materials. Second, these developments foster the use of abstract machines that are systematically inferred from data, rather than being designed in an ad~hoc manner, as~has more typically been the~case.

To achieve the aim of this paper, we provide a pedagogical explanation of the application of computational mechanics to the nn and nnn Ising models, along with the necessary background from statistical mechanics, measure theory, stochastic processes, and information theory. We then apply these techniques to a wider range of spin models, such as finite-range Ising models, solid-on-solid models, and~three-body models. In~parallel, we find that the typical patterns observed in these spin models at various parameter values match those predicted by information measures and $\epsilon$-machines. This allows us to present an account of spin patterns that is clearly consistent with statistical mechanics and information/computation~theory.


\section{\label{sec:background}Background and~Methods}

This section provides an intuition-first, pedagogical introduction to the concepts and methods required for our information- and computation-theoretic analysis of 1D spin lattice models. While the underlying machinery is standard in the literature~\cite{feldman1998dnco, shalizi1999thermo, shalizi2001compmech}, we restrict attention to the ingredients strictly necessary for our purposes, motivate each formal object intuitively, and~ground each one in statistical mechanics. We provide detailed explanations of what each mathematical component means and what role it plays, instead of developing an abstract-first treatment aimed at greater formal~generality.

\subsection{\label{sec:spin_measurements}Spin Measurements: Boltzmann Distribution of Finite Chain Embedded in Infinite~Chain}

\begin{figure}[H]

    \begin{minipage}{0.6\textwidth}
        
        \includegraphics[width=\linewidth]{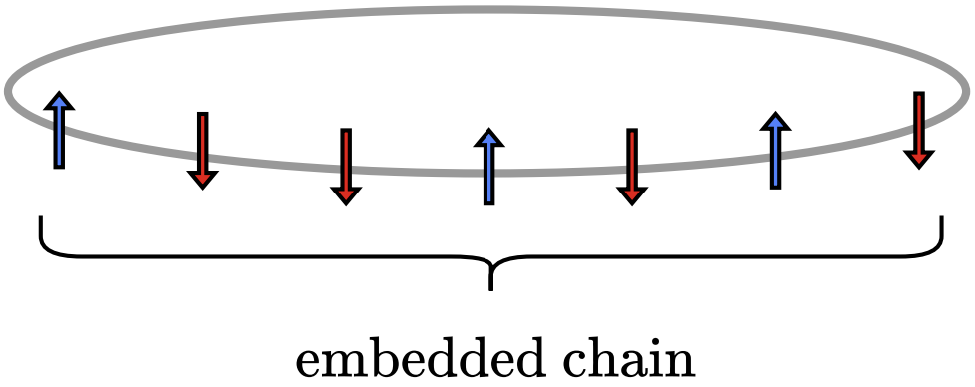}
    \end{minipage}
    \\[1.0em]

        \caption{Depiction 
 of a finite spin configuration embedded within an infinite spin configuration with periodic boundary~conditions.}\label{fig:embedded_chain}
   
\end{figure}

The Boltzmann distribution serves as an entry point for probing the structure of spin models; however, defining it for both finite and infinite configurations introduces significant difficulties. For~finite configurations, the~Boltzmann distribution lacks generality and often relies on numerical simulations for approximation~\cite{alves1990montecarlo, lin2013fpga, ferrenberg2018mclimits}. For~infinite configurations, a~different issue arises: their probability is zero~\cite{mackay2003info} (pp. 94--97). This defies our expectation that they occur and results in an unnormalized total probability—a sum that is zero rather than one. To~balance the constructiveness of finite configurations with the generality of infinite ones, we examine a hybrid configuration: a finite spin configuration embedded in an infinite one~\cite{myshlyavtsev2001embeddedprob}. Figure~\ref{fig:embedded_chain} illustrates the finite configuration embedded within the infinite one. The~key equations leading to the embedded distribution are presented below, with~detailed derivations provided in Appendices \ref{appendix:joint_prob_of_infinite_chain}, \ref{appendix:eigevalue_decomposition_of_transfer_matrix}, \ref{appendix:fixed_bd_conds_partition_function} and \ref{appendix:joint_prob_of_embedded_finite_chain}.

Consider a configuration consisting of $N$ spins, where each spin can take one of two values ($\uparrow$ or $\downarrow$) and interacts only with its nearest neighbors. For~convenience, the~configuration is subject to periodic boundary conditions:
\begin{equation}
s_0 \ldots s_{N-1} \ \text{where} \ s_0 = s_{N}
\label{eq:infinite_configuration}.
\end{equation}

The system is governed by a translationally-invariant Hamiltonian, that is, a~Hamiltonian whose form remains the same across spin sites. It is defined as follows:
\begin{equation}
E(s_i, s_{i+1}) = -Js_is_{i+1} - \dfrac{B}{2}(s_i+s_{i+1}).
\label{eq:translational_hamiltonian}
\end{equation}

Next, the~corresponding transfer matrix, with~components 
$V(s_i,s_{i+1})=e^{-\beta E(s_i s_{i+1})}$, is expressed as~\cite{yeomans1992} (p. 68):
\begin{equation}
\mathbf{V} = \begin{bmatrix}
    e^{-\beta E(\uparrow, \uparrow)} & e^{-\beta E(\downarrow, \uparrow)} \\
    e^{-\beta E(\uparrow, \downarrow)} & e^{-\beta E(\downarrow, \downarrow)} \\
    \end{bmatrix}
\label{eq:transfer_matrix}
\end{equation}

Then, the~probability distribution for this spin configuration in the thermodynamic limit $N \to \infty$ is obtained in terms of the transfer matrix components and the transfer matrix's principal eigenvalue $\lambda$ \cite{yeomans1992} (pp. 68--69):
\begin{equation}
\mathrm{Pr}(s_0, \ldots s_{N-1}) = \dfrac{\prod\limits_{i=0}^{N-1} V(s_i,s_{i+1})}{\lambda^N}
\label{eq:boltzmann_dist_infinite}
\end{equation}

Now, consider a specific finite configuration of length $L$ embedded in an infinite one:
\begin{equation}
\overrightarrow{s} = s_{0} \ldots s_{L-1} \ \text{where} \ L<N
\label{eq:embedded_configuration}
\end{equation}

Although the principal eigenvectors of the transfer matrix are seldom calculated in studies of spin models, they play a crucial role in defining the embedded distribution. Therefore, we obtain the normalized principal left and right eigenvectors of the transfer matrix, as~provided in Ref.~\cite{yeomans1992} (pp. 72--73). For~conciseness, these are expressed in terms of the magnetization $m$, as~shown below:
\begin{equation}
\textbf{u}^{\mathcal{L}} = \begin{bmatrix}
    \sqrt{\dfrac{1+m}{2}} & \sqrt{\dfrac{1-m}{2}} \\
    \end{bmatrix} \ \text{and} \ \textbf{u}^{\mathcal{R}} = \begin{bmatrix}
    \sqrt{\dfrac{1+m}{2}} \\
    \sqrt{\dfrac{1-m}{2}} \\
    \end{bmatrix}
\label{eq:ising_right_left_eigenvectors}
\end{equation}

Note that for the nn Ising model, the left and right eigenvectors are identical. Hence, in~all subsequent subsections of this section, we omit the left and right~superscripts.

Lastly, the~probability distribution for the embedded configuration~\cite{myshlyavtsev2001embeddedprob} is given by the following:
\begin{equation}
\mathrm{Pr}(\overrightarrow{s}) = \dfrac{u^{\mathcal{L}}_{s_{0}} u^{\mathcal{R}}_{s_{L-1}}\prod\limits_{i=0}^{L-2} V(s_i, s_{i+1})}{\lambda^{L-1}}
\label{eq:boltzmann_dist_embedded}
\end{equation}

Here, we provide the physical interpretation for each part of the equation:

\begin{itemize}
    \item In the denominator, $\lambda$ is raised to $L-1$ as each embedded configuration has $L$ spins and its boundaries are not periodic. 
    \item In the numerator, the~product of transfer matrix components consists of $L-1$ factors. This reflects the fact that only the spins within the bulk have neighboring spins to interact with on both their left and right sides.
    \item Also in the numerator, we include two extra terms: $u^{\mathcal{L}}_{s_{0}}$ and $u^{\mathcal{R}}_{s_{L-1}}$, which are the normalized principal eigenvector components associated with the boundary spins $s_{0}$ and $s_{L-1}$. Since the embedded configuration does not have periodic boundaries, these extra terms ensure that the boundary spins contribute to the system's magnetization as much as the bulk spins. Moreover, these terms are key to normalizing the joint probabilities.
\end{itemize}

To facilitate later discussion, it will be useful to denote the component associated with spins $\uparrow$ or $\downarrow$ as $u_{\uparrow}$, and~$u_{\downarrow}$, respectively.  The~values of these components correspond to either $u^{\mathcal{L}}_{s_{0}}$ or $u^{\mathcal{R}}_{s_{L-1}}$, depending on whether the orientations of the spins $s_0$ and $s_{-1}$ are up or down. For~example, in~a spin configuration like $\downarrow \uparrow \uparrow \uparrow$, the~component for the first spin $s_0$ is $u^{\mathcal{L}}_{s_{0}} = u_{\downarrow} = \sqrt{\frac{1-m}{2}}$, while the component for the last spin $s_3$ is $u^{\mathcal{R}}_{s_{3}} = u_{\uparrow} = \sqrt{\frac{1+m}{2}}$.

Alternatively, Equation~\eqref{eq:boltzmann_dist_embedded} can be interpreted as the probability measure of a coarse-grained configuration. The~nature of this coarse-graining and its implementation, which relies on measure theory, will be discussed in the following~section.

\subsection{\label{sec:measure_theory}Coarse-Graining via Measure~Theory}

In this section, we view finite configurations embedded in infinite ones as coarse-grained versions of infinite-spin configurations. Here, ``coarse-grained'' means a simplified representation that retains essential features while reducing detail~\cite{Flack2017}. The~procedure for arriving at these representations—that is, coarse-graining—is up to the scientist's discretion~\cite{shalizi2008macrostate}. However, when treating the spin model as a stochastic process, the~conventional approach is to reduce the degrees of freedom such that only contiguous ones remain~\cite{LeNy2007contiguousspins, muir2011gibbsmeasures, ganikhodjaev2007gibbsmeasure}. This coarse-graining is physically motivated by the observer's inability to record infinite measurements or degrees of freedom. To~define the set of coarse-grained configurations mathematically, we begin with the full set of possible~configurations.

Consider the set of all possible infinite spin configurations $\Omega$. An~individual configuration in this set is represented as $ \sigma \in \Omega $. The~degree of freedom at a lattice site $i$ within a configuration $\sigma$ is denoted by $\sigma_i$. Thus, a~configuration in terms of its degrees of freedom is given~by the following:

\begin{equation*}
    \sigma = \sigma_{0} \ldots \sigma_{N-1}
\end{equation*}
with
\begin{equation*}
    \sigma_{0} = \sigma_{N} \ \mbox{and} \ N \to \infty
\end{equation*}

The set of coarse-grained configurations $\Omega_C$ is defined as the set of infinite configurations in which the contiguous spins from $\sigma_{0}$ to $\sigma_{L-1}$ have fixed indices and can take any value from $\{-1, 1\}$.

This can be expressed as follows:
\begin{equation*}
    \Omega_C = \{\sigma \in \Omega \mid \sigma_{0}, \ldots, \sigma_{L-1} \ \text{have fixed indices} \}.
\end{equation*}

Alternatively, the~set of coarse-grained configurations can be defined as follows:
\begin{equation*}
    \Omega_C = \{ C_1, C_2, \ldots \}
\end{equation*}

with each coarse-grained configuration $C_j$ defined as follows:
\begin{equation*}
    C_j = \{ \sigma \in \Omega \mid \sigma_{0} = s_0, \ldots, \sigma_{L-1} = s_{L-1} \}
\end{equation*}
where $s_0, \ldots, s_{L-1}$ represent the fixed spin values at fixed indices $0, \ldots, L-1$. In~more compact notation, this is written as follows:
\begin{equation*}
    C_j = \{\sigma \in \Omega \mid \sigma^L = s^L \}
\end{equation*}

\begin{figure}[H]

\begin{minipage}{0.7\textwidth}
    
    \includegraphics[width=\linewidth]{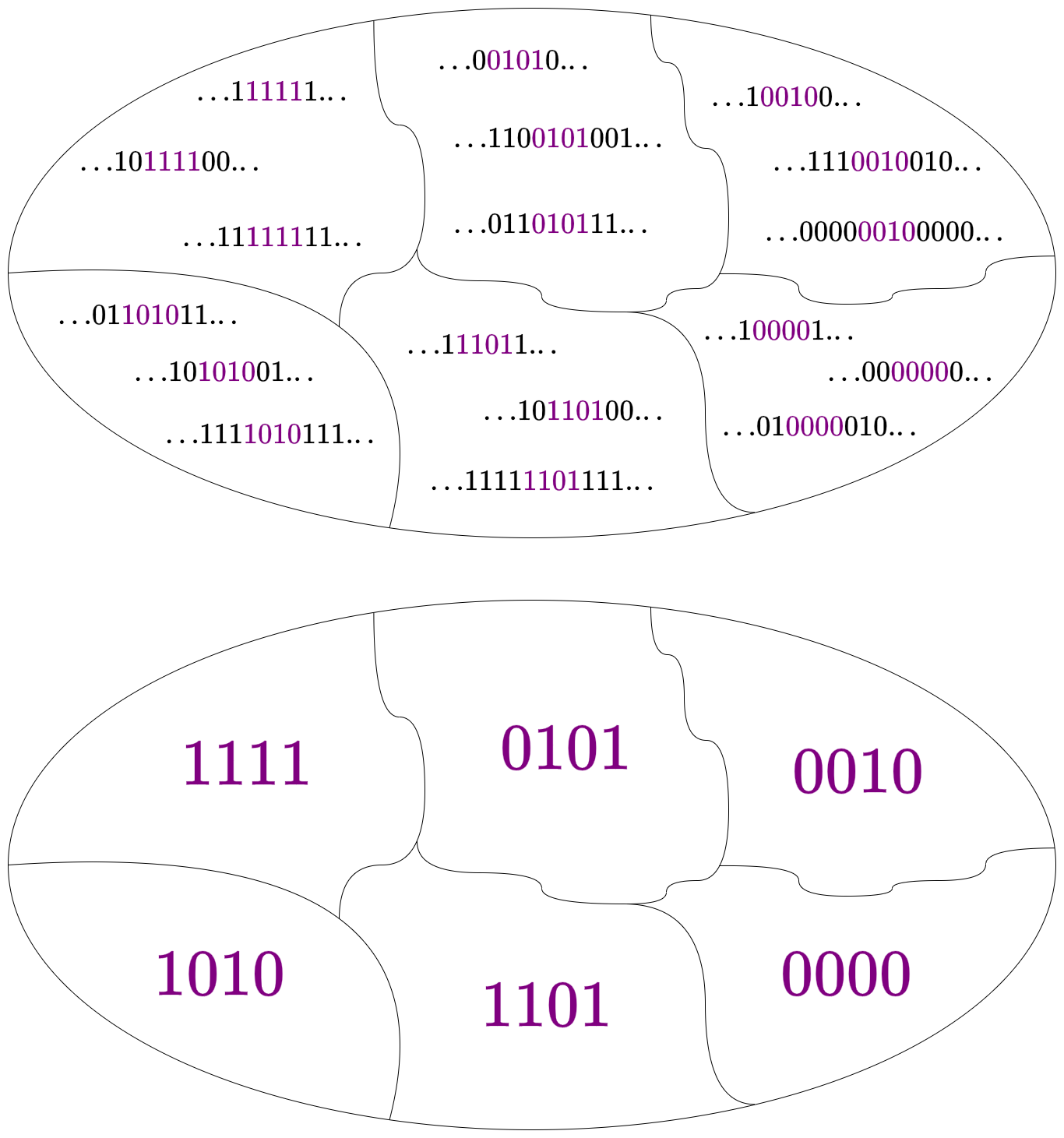}
\end{minipage}

    \caption{Graphical representation of coarse–grained Ising phase space. Only the purple spins are assigned fixed indices. For~clarity, down spins 
 $\downarrow$ are represented as $0$ instead of $-1$.} \label{fig:coarse–graining_ising_phase_space}

\end{figure}

Notably, the~act of coarse-graining changes our focus from individual configurations to sets, where each set $C_j$ groups configurations by their shared spin values. Figure~\ref{fig:coarse–graining_ising_phase_space}, shows how the set of all possible infinite spin configurations $\Omega$ is partitioned into the set of coarse-grained configurations $\Omega_C$. Accordingly, we must adapt our notion of probability to align with this perspective, transitioning from the concept of a probability distribution to that of a probability measure, as denoted by $\mu$ \cite{lind2021symbolic} (pp. 331--336).

To formalize this, we introduce the concept of a sigma algebra, denoted by $\mathcal{A}$.  This is a collection of all subsets of $\Omega_C$ that can be consistently assigned probabilities or measured, meaning they are physically~relevant.

The sigma algebra $\mathcal{A}$ has three key properties:

\begin{enumerate}
    \item \textbf{Entire Set Containment:
} $\mathcal{A}$ includes the sample space. In~this case, that is the coarse-grained set of all infinite configurations $\Omega_C$:
    \begin{equation*}
        \Omega_C \in \mathcal{A}
    \end{equation*}

    \item \textbf{Complement Closure:} If a set $A$ is in $\mathcal{A}$, then its complement $\Omega_C \setminus A$ must also be in~$\mathcal{A}$:
    \begin{equation*}
        A \in \mathcal{A} \implies \Omega_C \setminus A \in \mathcal{A}
    \end{equation*}

    \item \textbf{Countable Union Closure:} If $A_1, A_2, A_3, \dots$ are in $\mathcal{A}$, then their countable union is also in $\mathcal{A}$:
    \begin{equation*}
        A_1, A_2, \dots \in \mathcal{A} \implies \bigcup_{i=1}^{\infty} A_i \in \mathcal{A}
    \end{equation*}
\end{enumerate}

With the concept of a sigma algebra established, we can now turn to the probability measure. This measure is analogous to a probability distribution, but~applies to sets rather than individual outcomes. It extends the key constructive properties of probability distributions—namely, nonnegativity, normalization, and additivity—from finite to infinite~configurations. 

The probability measure is formalized as a function
\begin{equation*}       
    \mu: \mathcal{A} \rightarrow [0, 1],
\end{equation*}
which assigns a probability to each event in $\mathcal{A}$ and satisfies the following three key~properties:
\begin{enumerate}
    \item \textbf{Nonnegativity:} 
    In the same way that joint probabilities for finite configurations are never negative, the~probability measure assigned to any set in $\mathcal{A}$ must also be~\mbox{nonnegative}.

    \begin{equation*}
        \mu(A) \geq 0 \text{ for every } A \in \mathcal{A}.
    \end{equation*}
    \item \textbf{Normalization:} Similar to the sum of joint probabilities for all configurations equaling 1, the~probability measure for the entire sample space, the~set of coarse-grained configurations $\Omega_C$, must be $1$.
    \begin{equation*}
        \mu(\Omega_C) = 1 
    \end{equation*}
    \item \textbf{Countable additivity:} Mirroring the additivity of joint probabilities, which asserts that the total probability of finite configurations equals the sum of their individual probabilities, probability measures demonstrate countable additivity. This property dictates that for any countable collection of non-overlapping sets (cylinder sets) $\{A_i\}_{i=1}^{\infty}$, the~probability of their union is the sum of the probabilities of the individual sets:
    \begin{equation*}
        \mu\left(\bigcup_{i=1}^{\infty} A_i\right) = \sum_{i=1}^{\infty} \mu(A_i),
    \end{equation*}
    where each $A_i$ is a cylinder set corresponding to a coarse-grained configuration, and~the union represents the combined event of these~configurations. 

\end{enumerate}

The last step in constructing the spin probability measure involves assigning each spin cylinder set's probability measure the value of its associated embedded configuration's probability. Notably, information measures in later sections are denoted with a $\mu$ subscript, indicating that their argument is a probability measure~\cite{feldman1998dnco}.

\subsection{\label{sec:stochastic_processes}System and Measurements: Stochastic~Processes}

As mentioned in the introduction, we interpret configurations as realizations of a stochastic process.  This section aims to delve further into this formalism by first explaining the reasons for departing from the conventional~approach.

Traditionally, a~spin configuration is represented as an event $s$ of a random variable $S$. For~example, a~configuration with all spins pointing up is depicted as follows:
\begin{equation}
s = \ldots \uparrow \uparrow \uparrow \ldots
\label{eq:all_ups_config_as_event}
\end{equation}

However, this formalism impedes a direct examination of individual spins and their interactions. Furthermore, it leads to an unwieldy number of possible events. To~address these issues, we adopt a more nuanced approach. Instead of representing a configuration as a single event, we depict it as a specific realization of events:
\begin{equation}
\overleftrightarrow{s} = \ldots s_{-1} s_0 s_1 \ldots
\label{eq:specific_realization}
\end{equation}

This realization is an instance of a stochastic process, i.e.,~a partially-ordered chain of random variables:
\begin{equation}
\overleftrightarrow{S} = \ldots S_{-1} S_0 S_1 \ldots
\label{eq:stochastic_process}
\end{equation}
whose associated probability distribution is given by the following:
\begin{equation}
\operatorname{Pr}\left(\ldots S_{-1} S_{0} S_{1} \ldots \right)
\label{eq:stochastic_process_dist}.
\end{equation}

Within this framework, the~all-ups spin configuration is now denoted as follows:
\begin{equation}
\overleftrightarrow{s} = \ldots \uparrow \uparrow \uparrow \uparrow \ldots
\label{eq:all_ups_config_as_realization}
\end{equation}

Without loss of generality, we can split a process into two parts: the past process, defined as follows:
\begin{equation}
\overleftarrow{S} = \ldots S_{-1}
\label{eq:all_ups_config_as_realization_2}.
\end{equation}
along with its associated past realization, and~the future process, defined as follows:
\begin{equation}
\overrightarrow{S} = S_{0} \ldots
\label{eq:all_ups_config_as_realization_3}
\end{equation}
along with its associated future~realization.

For simplicity, we will use the terms ``past'' and ``future'' to refer to both processes and their associated realizations, with~the specific meaning inferred from the~context.

The spin stochastic process will be our object of study. In~the following subsection, we will elaborate on how it relates to broader categories of processes, as~seen in Ref.~\cite{feldman_2001_regularities_unseen}.

\subsubsection{\label{sec:types_of_processes}Types of~Processes}

\paragraph{Stationary Process}
A process in which the statistical properties of its random variables remain invariant over time. These properties include but are not limited to mean, variance, or~joint distribution.
\paragraph{Strictly Stationary Process}
A process whose joint distribution remains invariant under shifts in time. In~other words, a~process whose random variables are time-translation invariant. That is, a~process that satisfies the following:
\begin{equation}
\operatorname{Pr}\left(S_t S_{t+1} \ldots S_{t+L-1}\right) = \operatorname{Pr}\left(S_0 S_1 \ldots S_{L-1}\right).
\label{eq:strictly_stationary_process}
\end{equation}

\paragraph{Markovian Process}

A process in which the probability distribution of the next random variable depends only on the preceding one. That is, a~process whose joint distribution factors as follows:
\begin{equation}
\operatorname{Pr}(\stackrel{\leftrightarrow}{S})=\ldots \operatorname{Pr}\left(S_{i} \mid S_{i-1}\right) \operatorname{Pr}\left(S_{i+1} \mid S_{i}\right)\ldots
\label{eq:markovian_process}
\end{equation}

\paragraph{R-Order Markovian Process}

A process in which the probability distribution of the next random variable depends only on the $R$ preceding ones. That is, a~process whose joint distribution is given as follows:
\begin{equation}
\operatorname{Pr}(\stackrel{\leftrightarrow}{S})=\ldots \operatorname{Pr}\left(S_{i} \mid S_{i-R}, \ldots, S_{i-1} \right) \ldots
\label{eq:r_order_markovian_process}
\end{equation}

\paragraph{Spin Process}
A process whose associated probability distribution is generated by a spin Hamiltonian model. For~the models considered in this work (finite-range Ising, Solid on solid, and Three-body models), this process is strictly stationary and Markovian or $R$-order~Markovian. 

We can now define information measures of randomness, regularity, and structure for a stochastic process, starting from the basics of information~theory.

\subsection{\label{sec:info_measures}Information~Measures} 

What is information? Information can be conceived as quantifiable surprise, defined in terms of probabilities~\cite{tribus_thermostats_thermo} (p. 64). Through this lens, an~event $s$ that is not likely to occur is deemed surprising, thus carrying high informational content. This means that the information of an event $H(s)$ is inversely proportional to its probability, that is, $H(s) \propto \dfrac{1}{p(s)}$. More specifically, the~event's information content—termed self-information—\cite{tribus_thermostats_thermo} (p. 64) is defined as follows:
\begin{equation}
H(s) = -\log_2 p(s)
\label{eq:self_informaiton}.
\end{equation}

Here, the~presence of the logarithm is a convenient guarantee that the self-information possesses the additive property~\cite{shannon1963math}. That is, the~total surprise from combining events $1$ and $2$ equals the sum of their individual~surprises.

The natural next step is to consider a random variable $S$. Its information content is known as Shannon entropy. It is defined as the weighted sum of the self-information of each possible event within the variable. Mathematically, it is expressed as follows:
\begin{equation}
H(S) = -\sum\limits_{s=\pm 1} \log_2 p(s)
\label{eq:shannon_entropy}
\end{equation}

Following this line of reasoning, we can define the conditional entropy (Ref.~\cite{shannon1963math}; Ref.~\cite{cover2006info}, {p. 17}
) as the amount of information needed to specify a random variable $S_1$ given that a random variable $S_0$ is known.
\begin{equation}
H(S_1|S_{0})  = -\sum\limits_{s_0, s_1 = \pm 1} \mathrm{Pr}(s_0,s_1)\log_2\mathrm{Pr}(s_1|s_0)
\label{eq:conditional_entropy}
\end{equation}

Moreover, we can define the joint entropy (Ref.~\cite{shannon1963math}; {Ref.}~\cite{cover2006info}, {pp. 16--17}) as the amount of information contained in two random variables.
\begin{equation}
H(S_0, S_1) = -\sum\limits_{s_0, s_1 = \pm 1} \mathrm{Pr}(s_0, s_1) \log_2 \mathrm{Pr}(s_0, s_1)
\label{eq:joint_entropy}
\end{equation}

Now, how may we define the entropy of our object of interest, that is, the~stochastic process? The simplest answer would be to consider the growth entropy~\cite{feldman1998infotutorial}, that is, the~Shannon entropy of the entire process.
\begin{equation}
H(S^L) = -\sum\limits_{s_0 = \pm 1} \ldots \sum\limits_{s_{L-1} = \pm 1} \mathrm{Pr}(s^L) \log_2 \mathrm{Pr}(s^L)
\label{eq:growth_entropy}
\end{equation}

However, as~the length of the process increases, the~growth entropy also rises and ultimately diverges when the process extends towards infinity ($L \to \infty$). This raises the question: how can we capture the total information of a stochastic process? A solution lies in the Shannon entropy rate (Ref.~\cite{feldman1998infotutorial}; Ref.~\cite{cover2006info}, pp. 74--76) defined as follows:
\begin{equation}
h_\mu = \lim\limits_{L \to \infty} \dfrac{H(S^L)}{L}
\label{eq:entropy_rate}
\end{equation}

Again, the~symbol $\mu$ signifies that the Shannon entropy rate is calculated in terms of a probability measure. Notably, this rate can be simplified for stationary, Markovian processes, such as the spin process. For~a stationary process, the~entropy rate reduces to the following:
\begin{equation}
h_\mu = H(S_L|S_{L-1}, \ldots, S_1)
\label{eq:entropy_rate_for_stationary}.
\end{equation}

If the process is also Markovian, it becomes the following:
\begin{equation}
h_\mu = H(S_0|S_{-1})
\label{eq:entropy_rate_for_stationary_and_markovian}
\end{equation}

By recasting the Shannon entropy rate as a conditional entropy, we can understand it as the amount of surprise each spin contributes. This effectively measures the process's randomness per spin. Furthermore, for~one-dimensional spin models, the~Shannon entropy rate matches the Boltzmann entropy density, the~more familiar form of entropy in statistical mechanics, as~shown in Appendix \ref{appendix:hmu_versus_htherm}.

Since the regularity of the spin process is interpreted as the information shared between the process' past and future, the~regularity is defined as the process' mutual information or excess entropy~\cite{shaw1984, crutchfield1983, grassberger1986, lindgren1988}. Mathematically, it is defined for the spin process as follows:
\begin{equation}
\mathbf{E} = I(\overleftarrow{S};\overrightarrow{S}) = I(S_{-1};S_0)
\label{eq:excess_entropy}.
\end{equation}

Therefore,
\begin{equation}
\begin{aligned}
\mathbf{E} = \sum_{s_{-1}, s_0 = \pm 1} \mathrm{Pr}(s_{-1},s_0) \log_2 \left(\frac{\mathrm{Pr}(s_{-1},s_0)}{\mathrm{Pr}(s_{-1})\mathrm{Pr}(s_0)}\right)
\end{aligned}
\label{eq:excess_entropy_in_terms_of_probs}.
\end{equation}

Notably, the~excess entropy 
$\mathbf{E}$ can be interpreted as \textit{predictable} information. That is, it quantifies the amount of information an observer has for recognizing configuration patterns, even if that information is not enough to identify them. However, in~the absence of entropy rate $h_{\mu}$, $\textbf{E}$ is sufficient to determine how much information is required for the observer to achieve \textit{synchronization} with the underlying configuration patterns. Synchronization, from~this purely information-theoretic perspective, refers to the observer's ability to recognize and discern these configuration~patterns.

To measure the process's structure or statistical complexity, we need to determine the asymptotic probabilities of its patterns or causal states $\mathcal{S}$. In~general, this often requires inferring the process's $\epsilon$-machine, especially for non-Markovian processes~\cite{shalizi2001phdthesis} (p. 37).  However, for~the spin process, we can calculate them directly since we have a natural definition of causal~states. 

Given the Markovian nature of the spin process, the~next spin only depends on the previous one. Thus, the~probability distribution of future spins conditioned on past ones matches the probability distribution of the future conditioned on the previous spin being up or down. Now, since the probability of a spin up and the probability of a spin down add up to 1, and~they represent the probability per site throughout the process, they can be interpreted as the asymptotic probabilities of the causal states. Therefore, the~statistical complexity of the spin process can be quantified as follows~\cite{feldman1998dnco}:
\begin{equation}
C_{\mu} = H(\text{``patterns''}) = H(\mathcal{S}) = H(S_0)
\label{eq:stat_complexity}.
\end{equation}

Therefore,
\begin{equation}
\begin{aligned}
C_\mu &= - \sum_{s_0=\pm1} \mathrm{Pr}(s_0) \log_2 \mathrm{Pr}(s_0) \\
&= - \sum_{s_0=\pm1} (u_{s_0}^L u_{s_0}^R) \log_2 (u_{s_0}^L u_{s_0}^R) \\
&= - \sum_{s_0=\pm1} u_{s_0}^2 \log_2 u_{s_0}^2
\end{aligned}
\label{eq:stat_complexity_derivation}.
\end{equation}

This information can be simply related via the identity $H(S_0) = H(S_0|S_{-1}) + I(S_{-1}, S_0)$ as
\begin{equation}
C_{\mu} = R h_{\mu} + \mathbf{E}
\label{eq:info_measures_identity}.
\end{equation}

This relationship~\cite{shalizi2001phdthesis} (p. 37) formalizes our intuition that structure is a blending of randomness and regularity. Here, $R$ denotes the neighborhood radius, which equals 1 for the nn Ising~model.

Since the causal states are sufficient to predict the process's future, and~considering that prediction is tantamount to reproduction, statistical complexity can be defined as the minimum amount of information required to reproduce the stochastic process~\cite{feldman1998dnco}. As~mentioned in the introduction, if~structure is viewed as quantifiable information, then this suggests that the mechanism generating the structure can be described as a machine~\mbox{\cite{shalizi1999thermo, crutchfield2012btworderchaos}.}

\subsection{\label{subsec:comp_mech}Structure: Computational~Mechanics}

To formalize the mechanism generating a physical process's structure, the~concept of a machine must be adapted to satisfy three statistical mechanical~constraints:
\begin{enumerate}
    \item Be capable of reproducing ensembles;
    \item Possess a well-defined notion of ``state'';
    \item Be derivable from first principles.
\end{enumerate}

Computational mechanics meets the first requirement by enhancing the simplest machine in TOC, the~Deterministic Finite State Machine (DFSM), with~probabilistic features while keeping its determinism intact~\cite{shalizi2001compmech, marzen2022pdfr}. The~former is achieved by incorporating probabilities into the state transitions, and~the latter is maintained by ensuring that the probability of transitioning to the next state, given the current state and a specific outgoing symbol, is precisely one. These modifications result in a machine known as Probabilistic Finite State Machine (PDFM) or $\epsilon$-machine.

The second requirement is fulfilled by operationalizing TOC's conceptual definition of a state—an entity that ``remembers a relevant portion of the system's history''~\cite{hopcroft2001} (pp.~2--3)—as a causal state. A~causal state is the collection of all past realizations that when individually conditioning the process' future yield the same conditional probability distribution~\cite{shalizi2001compmech, crutchfield2012btworderchaos}. Notably, formalizing the notion of ``state'' is crucial not just for conceptual clarity, but~also to satisfy the third requirement. The~reason for this is that without a clear understanding of what states are, the~procedure for inferring them is much less~clear.

To satisfy the third condition, we recast the definition of causal state as a guiding principle for inferring causal states from realizations, that is, the~causal equivalence principle~\cite{shalizi2001compmech, crutchfield2012btworderchaos}. It states that two past realizations belong to the same causal state if they yield the same conditional distributions over the process's futures. In~practice, this principle allows us to construct the underlying $\epsilon$-machine of an~ensemble.

In summary, the~key ingredients of computational mechanics are the concepts of $\epsilon$-machine, causal transition, causal state, and~the causal equivalence principle~\cite{shalizi2001compmech}.  While we introduced them in this order to capture how they would be rediscovered conceptually, we will now present them in reverse order to delve into their mathematical details more~pedagogically.

\vspace{12pt}

\textbf{Causal equivalence principle.} Two pasts are considered causally equivalent if and only if they make the same prediction over the future, i.e.,
\begin{equation}
\stackrel{\leftarrow}{s} \sim \stackrel{\leftarrow}{s}' \Longleftrightarrow \mathrm{Pr}(\stackrel{\rightarrow}{S}|\stackrel{\leftarrow}{s}) = \mathrm{Pr}(\stackrel{\rightarrow}{S}|\stackrel{\leftarrow}{s}')
\label{eq:causal_equivalence_principle}
\end{equation}

Effectively, this principle groups pasts that lead to the same future into what are known as causal states. To~formalize what we mean by ``leads,'' a causal state is defined~as follows:

\vspace{12pt}

\textbf{Causal state.} A triple that~contains the following:

\begin{enumerate}
    \item An event with its associated probability of the causal state random variable $\mathcal{S}$:
\begin{equation}
    \mathcal{S}_i \ \text{and} \ \mathrm{Pr}(\mathcal{S}_i)
    \label{eq:causal_event}.
    \end{equation}
    \item A distribution of the future conditioned on the causal event, i.e.,~a morph:
\begin{equation}
    \mathcal{M}_i = \mathrm{Pr}(\stackrel{\rightarrow}s|\mathcal{S}_i)
    \label{eq:morph}.
    \end{equation}
    \item The set of histories that lead to the same morph:
\begin{equation}
    \mathcal{H}_i = \{\stackrel{\leftarrow}s | \ \mathrm{Pr}(\stackrel{\rightarrow}{S}|\mathcal{S}_i) = \mathrm{Pr}(\stackrel{\rightarrow}{S}|\stackrel{\leftarrow}{s}) \}
    \label{eq:set_of_histories}.
    \end{equation}
\end{enumerate}

Now, assuming that our machine is deterministic in the computational theoretic sense, we can define the causal transition~as follows:

\vspace{12pt}

\textbf{Causal transition.} The probability of transitioning from state $\mathcal{S}_i$ to state $\mathcal{S}_j$ while emitting the symbol $s \in \mathcal{A}$:
\begin{equation}
\begin{aligned}
T^{(s)}_{ij} &= \mathrm{Pr}(\mathcal{S}_j, s|\mathcal{S}_i) \\
             &= \mathrm{Pr}(\mathcal{S}_j|s,\mathcal{S}_i)\mathrm{Pr}(s|\mathcal{S}_i) \\
             &= \mathrm{Pr}(s|\mathcal{S}_i)
\end{aligned}
\label{eq:causal_transition}.
\end{equation}

These definitions allows us to construct the minimal machine supporting a stochastic process'~structure.

\vspace{12pt}

\textbf{\boldmath{$\epsilon$}-machine or PDFM.} A pair that~contains the following:
    \begin{enumerate}
        \item The set of causal states;
        \item Transition dynamic (causal transitions gathered in a matrix) ~\cite{crutchfield1994calculi}.
    \end{enumerate}

For inferring $\epsilon$-machines, it will be important to distinguish between two types of causal states:

\begin{itemize}
    \item \textbf{Recurrent causal states}: These are states to which the machine will repeatedly transition as it operates. Consequently, their asymptotic probability is~non-zero.
    
    \item \textbf{Transient causal states}: These are states that the machine may reach temporarily but will not return to. As~a result, their asymptotic probability is zero: $\mathrm{Pr}( \mathcal{S}_i)=0$.
\end{itemize}

Notably, the~connectivity and number of transient states specify how difficult it is to identify the periodicity of configurations. In~other words, these transient states reflect the computational effort required to achieve synchronization with the recurrent causal states. Here, synchronization is recast as the observer achieving certainty about the recurrent causal state it occupies, even in systems with nonzero entropy rate $h_{\mu}$. Thus, transient states offer a computational perspective on synchronization, which completes the informational interpretation provided by the excess entropy $\textbf{E}$.

Since this is a principled approach, we can infer our machine of interest, rather than design it. For~spin processes, an~analytical method exists for inferring recurrent causal states~\cite{feldman1998dnco}. Moreover, transient states can be reconstructed from these recurrent states, as~detailed in Appendix B of Ref.~\cite{feldman1998dnco}. Below, we provide a step-by-step explanation of the analytical reconstruction method for recurrent causal~states.

\subsubsection{\label{sec:analytical_epsilon}Analytical Method to Infer \texorpdfstring{$\epsilon$}{epsilon}-Machines}

\begin{enumerate}
    \item Consider a finite configuration of length $2L$ embedded in an infinite one. 
    $$\overleftrightarrow{s} = s_{-L} \ldots s_{-1} s_0 s_1 \ldots s_{L-1}$$   
    \item Consider the joint probability of the embedded finite configuration.
\begin{equation}
        \mathrm{Pr}(\overleftrightarrow{s}) = \dfrac{u^{\mathcal{L}}_{s_{-L}} u^{\mathcal{R}}_{s_{L-1}}\prod\limits_{i=-L}^{L-2} V_{s_is_{i+1}}}{\lambda^{2L-1}}
    \label{eq:full_embedded_configuration}
    \end{equation}
     
    \item Compute the conditional probability of the right half of the configuration given the left half. 
    $$\mathrm{Pr}(\stackrel{\rightarrow}{s} | \stackrel{\leftarrow}{s}) = \dfrac{u^{\mathcal{R}}_{s_{L-1}}}{u^{\mathcal{R}}_{s_{-1}}}\dfrac{\prod\limits_{i=-1}^{L-2} V_{s_is_{i+1}}}{\lambda^{L}}$$
    \item Notice that the only past element the conditional probability depends on is its last spin $s_{-1}$. Thus, the~conditional probability is Markovian. 
    $$\mathrm{Pr}(\stackrel{\rightarrow}{s} | \stackrel{\leftarrow}{s}) = \mathrm{Pr}(\stackrel{\rightarrow}{s} | \text{last spin})$$ 
    \item Identify morphs.
    $$\mathrm{Pr}(\stackrel{\rightarrow}{s} | \mathcal{S}_A) = \mathrm{Pr}(\stackrel{\rightarrow}{s} | \text{pasts whose last spin is} \ \uparrow)$$
    $$\mathrm{Pr}(\stackrel{\rightarrow}{s} | \mathcal{S}_B) = \mathrm{Pr}(\stackrel{\rightarrow}{s} | \text{pasts whose last spin is} \ \downarrow)$$
    \item Identify the number of causal~states. \\
    Since there are two morphs, there are two \\ causal states at most
    \item Identify sets of histories that lead to the same morph. 
    $$\{\stackrel{\leftarrow}s | \ \text{last spin is } \uparrow \} \ \text{and} \ \{\stackrel{\leftarrow}s | \ \text{last spin is } \downarrow \}$$ 
    \item Apply the definition of causal transitions.
    $$T^{(\uparrow)}_{\mathcal{A} \mathcal{A}} = \mathrm{Pr}(\uparrow|\uparrow) = \dfrac{e^{\beta (J+B)}}{\lambda}$$ $$T^{(\downarrow)}_{\mathcal{A} \mathcal{B}} = \mathrm{Pr}(\downarrow|\uparrow) = \dfrac{e^{-\beta J}}{\lambda} \sqrt{\dfrac{1-m}{1+m}}$$
    $$T^{(\downarrow)}_{\mathcal{B} \mathcal{B}} = \mathrm{Pr}(\downarrow|\downarrow) = \dfrac{e^{\beta (J-B)}}{\lambda}$$ $$T^{(\uparrow)}_{\mathcal{B} \mathcal{A}} = \mathrm{Pr}(\uparrow|\downarrow) = \dfrac{e^{-\beta J}}{\lambda} \sqrt{\dfrac{1+m}{1-m}}$$
    \item Calculate asymptotic causal state probabilities using two~facts:
    \begin{itemize}
        \item $\mathrm{Pr}(\stackrel{\rightarrow}{s} | \mathcal{S}_A) = \mathrm{Pr}(\stackrel{\rightarrow}{s} | \uparrow)$ 
        \item $\mathrm{Pr}(\stackrel{\rightarrow}{s} | \mathcal{S}_B) = \mathrm{Pr}(\stackrel{\rightarrow}{s} | \downarrow)$ 
        \item $\mathrm{Pr}( \mathcal{S}_A)+\mathrm{Pr}( \mathcal{S}_B)=1$
    \end{itemize}
    Since $\mathrm{Pr}( \uparrow)+\mathrm{Pr}( \downarrow)=1$, by~inspection, we~have the following:
    \begin{itemize}
        \item $\mathrm{Pr}( \mathcal{S}_A)=\mathrm{Pr}( \uparrow)=u^2_\uparrow=\dfrac{1+m}{2}$
        \item $\mathrm{Pr}( \mathcal{S}_B)=\mathrm{Pr}( \downarrow)=u^2_\downarrow=\dfrac{1-m}{2}$
    \end{itemize}
    \item 
    Build transition dynamic \textbf{T}.
    $$\textbf{T} = 
    \begin{bmatrix}
    0 & \mathrm{Pr}(\mathcal{S}_A) & \mathrm{Pr}(\mathcal{S}_B) \\
    0 & \mathrm{Pr}(\mathcal{S}_A|\mathcal{S}_A) & \mathrm{Pr}(\mathcal{S}_B|\mathcal{S}_A) \\
    0 & \mathrm{Pr}(\mathcal{S}_A|\mathcal{S}_B) & \mathrm{Pr}(\mathcal{S}_B|\mathcal{S}_B)
    \end{bmatrix}$$ 
    $$= \begin{bmatrix}
    0 & \frac{1+m}{2} & \frac{1-m}{2} \\
    0 & \frac{e^{\beta (J+B)}}{\lambda} & \frac{e^{-\beta J}}{\lambda} \sqrt{\frac{1-m}{1+m}} \\
    0 & \frac{e^{-\beta J}}{\lambda} \sqrt{\frac{1+m}{1-m}} & \frac{e^{\beta (J-B)}}{\lambda}
    \end{bmatrix}$$
    \item Find the left eigenvector using  $\langle \pi \vert \textbf{T} = \langle \pi \vert$.
    $$\langle \pi \vert = (0,\frac{1+m}{2},\frac{1-m}{2})$$ 
    Since $\textbf{T}$ is a stochastic matrix, this is its asymptotic probability distribution vector, which contains the causal states' probabilities, as~seen in Refs.~\cite{lind2021symbolic} (p. 330), \cite{young1994fluctuation} and 
 \cite{young1991grammar} (p. 128).
    \item Build HMM representation of $\epsilon$-machine using the transition matrix $\textbf{T}$. Details of the resulting machine, for~the parameter values $J_1 = 1.0$, $B = 0.35$, and~$T = 1.5$, are provided in  Appendix~\ref{appendix:nn_ising_machine}.
    
\end{enumerate}

\subsection{\label{sec:patterns}Patterns as \texorpdfstring{$\epsilon$-Machines}{epsilon-machines}}

\begin{figure}[H]

\begin{minipage}{0.9\textwidth}
    
    \includegraphics[width=\linewidth]{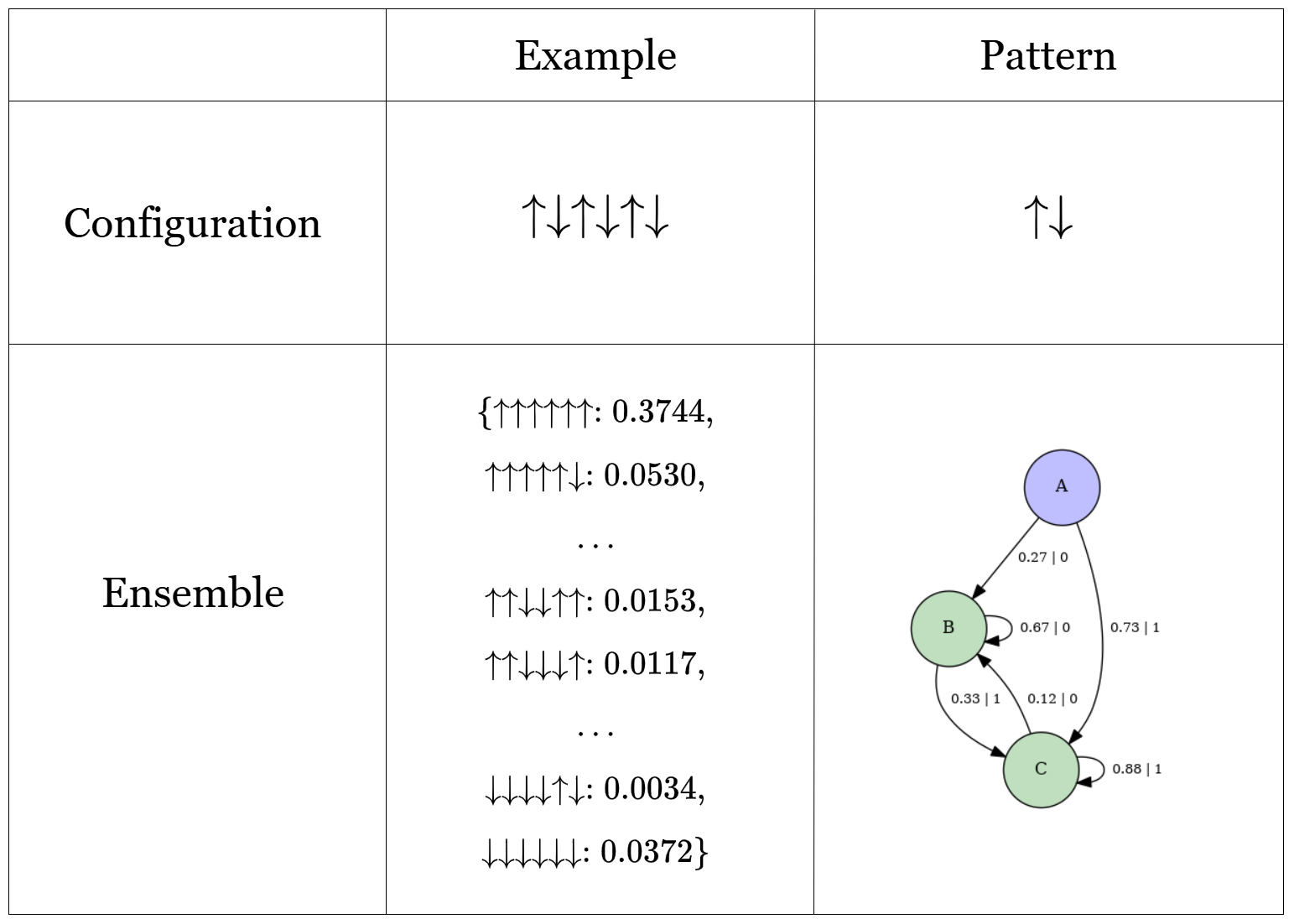}
\end{minipage}

    \caption{Graphical 
 representation of configuration and ensemble pattern~concepts.}
    \label{fig:config_vs_ensemble_patterns}

\end{figure}

The following example illustrates how computational mechanics formalizes the concept of a pattern. Consider a spin configuration such as $\uparrow \downarrow \uparrow \downarrow \uparrow \downarrow$. When asked, ``What's the pattern in this configuration?'', an~intuitive answer might be $\uparrow \downarrow$. However, if~presented with an ensemble of spin configurations and posed with the same question, the~concept of a pattern becomes vague. To~reason towards a definition of pattern for ensembles, we can ask: ``What's the key property of a pattern?''  A plausible candidate is that a pattern represents a compressed form of data that enables an observer to reproduce the original content~\cite{Dennett1991realpatterns}. Thus, we can then ask: ``What's the object that statistically reproduces such a configuration?'' The framework of computational mechanics provides the answer: The $\epsilon$-machine, which can be interpreted as a \textit{physical} or \textit{ensemble pattern} \cite{shalizi2001compmech, crutchfield2012btworderchaos}. Figure~\ref{fig:config_vs_ensemble_patterns} illustrates the relationship between configuration and ensemble~patterns. 

From this point forward, the~plotted machines will be derived using the CMPy package, which implements a tree-reconstruction method for inferring $\epsilon$-machines, as~described in Refs.~\cite{crutchfield1989inferring, crutchfield1994calculi}. The~transient and recurrent states of these machines are represented in purple and green, respectively. For~clarity in visualization, spins $\uparrow$ and $\downarrow$, emitted during transitions between causal states, are represented as $1$ and $0$, respectively. The~ensembles of spin models discussed in the following section include configurations of either $4$ or $6$ spins. Configurations with probabilities below $1 \times 10^{-5}$ are excluded from~consideration.

Based on Figure~\ref{fig:config_vs_ensemble_patterns}, one might be tempted to conclude that an $\epsilon$-machine is simply a Hidden Markov Model (HMM). However, that is not the case. The~difference stems from how states in HMM and $\epsilon$-machines are characterized; specifically, a~causal state in an $\epsilon$-machine is defined as a triple, as~mentioned earlier. In~contrast, the~conventional use of HMMs typically equates a state directly with the outcome of a random variable, treating the state as a singular entity rather than a triple. The~definition of a state in computational mechanics is crucial, as~it provides the foundation for inferring states from first principles  rather than manually designing them~\cite{crutchfield1989inferring}.

\section{Results and~Discussion}


As discussed in the previous section, the~embedded Boltzmann distribution generates a vast number of spin configurations, making the structure and pattern of an arbitrary configuration unrepresentative of the system's overall structure and patterns. However, to~compare the information measures and $\epsilon$-machines to the Boltzmann ensemble, it may still be useful to examine the structure and patterns of the individual configurations that are most optimally representative. To~achieve this, we focus on a specific kind of configuration: \textit{typical configurations}. These configurations are the most likely outcomes generated by the embedded Boltzmann distribution of a given spin model. Among~these, \textit{likely typical configurations} have probabilities that are significantly higher than those of non-typical configurations, whereas \textit{unlikely typical configurations} have probabilities that are only slightly higher than those of non-typical configurations. The~patterns present in these typical configurations are referred to as \textit{typical configuration patterns}.

The patterns and structures of both typical and non-typical configurations across different spin models are shaped by various parameters~\cite{kikuchi1955variousparams}. To~identify commonalities in how these parameters contribute to the configurations' structure and patterns, we propose classifying them into three distinct types. To~illustrate this, we will reference the nearest-neighbor (nn) Ising model as an example while defining each type of~parameter.

\begin{enumerate}
    \item \textbf{Randomness Parameter:} This parameter governs the degree of randomness within the system. As~it increases, it leads configurations to become more uniformly likely.  In~the nn Ising model, temperature $T$ usually fulfills this role. 
    \item \textbf{Periodicity Parameter (Type 1):} This parameter enhances periodicity and, as~it varies, biases the system toward configurations that consist exclusively of a single period. In~the nn Ising model, the~coupling constant $B$ exemplifies this. It induces period 1 configurations whether $B$ is significantly positive or negative. Specifically, a~high positive $B$ biases all spins to point upwards, while a high negative $B$ results in all spins pointing downwards.
    \item \textbf{Periodicity Parameter (Type 2):} Similarly, this parameter enhances periodicity but, as~it varies, steers the system towards typical configurations with multiple distinct periods. In~the nn Ising model, this role is played by the coupling constant $J$. A~high positive $J$ value tends to produce period 1 configurations (all spins up), akin to $B$, but~a negative $J$ value leads to alternating spin configurations (e.g., up-down-up-down), indicating that the typical configuration can be of period 2.
\end{enumerate}

\subsection{Finite-Range Ising~Model}

\begin{figure}[H]

\begin{minipage}{0.85\textwidth}
    
    \includegraphics[width=\linewidth]{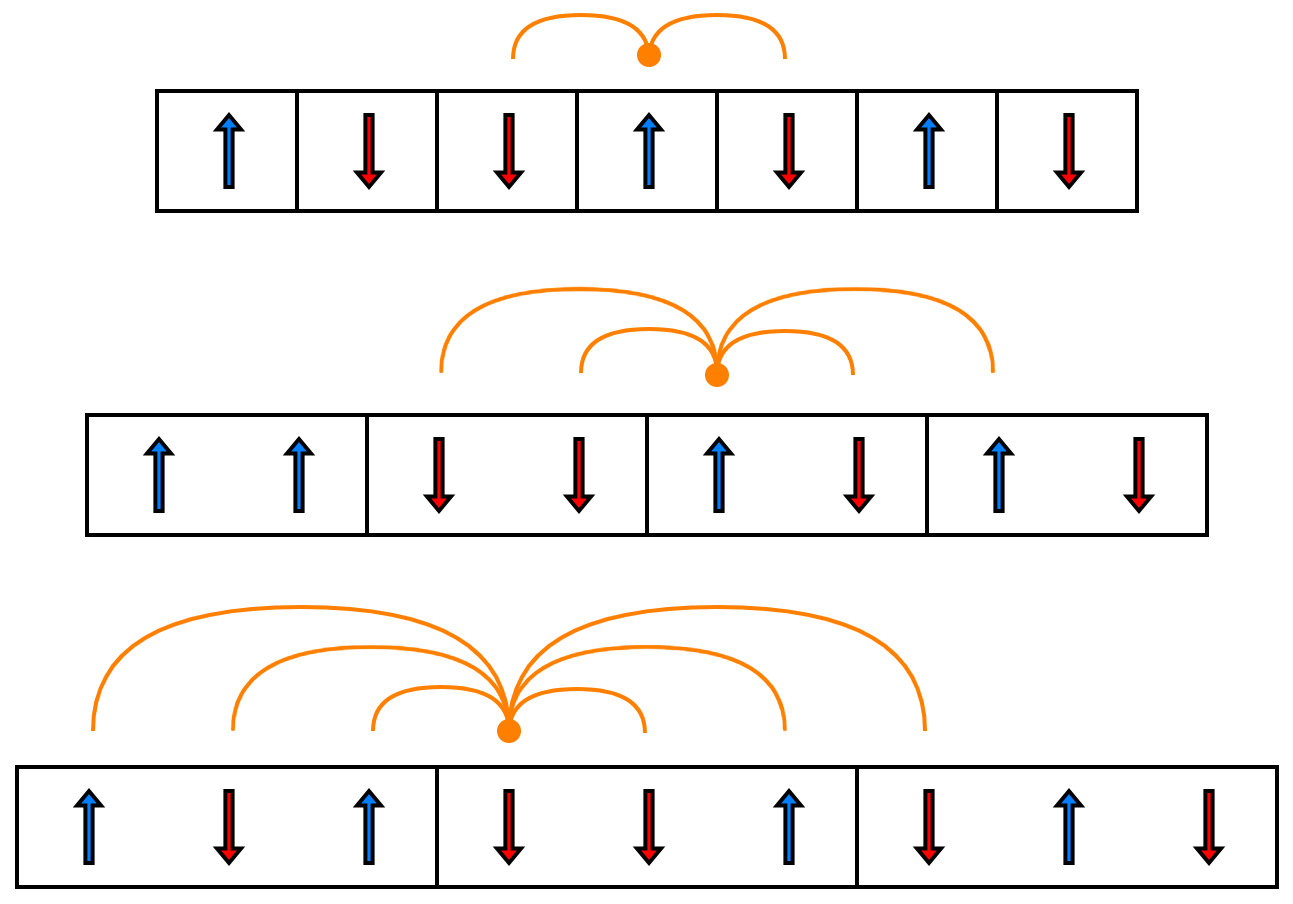}
\end{minipage}

    \caption{Illustration 
 of spin interactions in Ising models with neighboring radii $R = 1$ (\textbf{top}), \linebreak  $R = 2$ (\textbf{middle}), and~$R = 3$ (\textbf{bottom}).}\label{fig:finite_range_ising_model}

\end{figure}

The nearest-neighbor Ising model can be generalized to a finite-range model using Dobson's spin block method~\cite{dobson1969}. This approach consists of redefining the model's degrees of freedom from individual spins to blocks of spins. These spin blocks are only allowed to interact with their nearest-neighbor blocks. Equivalently, in~terms of spin variables, a~spin $s_i$ within a spin block $\eta_j$ is only allowed to interact with spins within the same block and spins within the nearest-neighbor spin blocks. Notably, every spin within a block will interact with all the spins within the same block. Nonetheless, a~given spin won't necessarily interact with all the spins from the nearest spin blocks  unless the nearest-neighbor Ising model is the specific model under consideration~\cite{dobson1969}. The~interactions of spins within spin blocks are illustrated in Figure~\ref{fig:finite_range_ising_model}. 

The spin block method expresses the Hamiltonian of two interacting spin blocks $\eta_j$ and $\eta_{j+1}$ of the finite-range Ising model as the sum of three contributions, shown in Equation~\eqref{eq:finite_range_ising_hamiltonian}. The~first is the energy within block $\eta_j$, encompassing the interactions among spins within the block as well as the interactions of each spin with the magnetic field. The~second contribution is the interaction energy between blocks $\eta_j$ and $\eta_{j+1}$, which is determined solely by the interactions between spins in $\eta_j$ and spins in $\eta_{j+1}$.  The~third contribution is the energy within block $\eta_{j+1}$, which, like the first, consists of the interactions between spins inside the block and the interactions of these spins with the magnetic field~\cite{dobson1969}. The~reduction of the finite-range Ising model Hamiltonian to the Hamiltonians of Ising models with neighboring radii $R = 1$, $2$, and~$3$ is shown in Appendix \ref{appendix:finite_range_ising_derivation}:
\begin{equation}
\begin{split}
E(\eta_j, \eta_{j+1}) = \frac{1}{2}X_{\eta_j} + Y_{\eta_{j}, \eta_{j+1}} + \frac{1}{2}X_{\eta_{j+1}}
\end{split}
\label{eq:finite_range_ising_hamiltonian}
\end{equation}
where
\begin{itemize}
    \item $X_{\eta_j} = -  \left( B\sum\limits_{i=0}^{n-1} s^{j}_i + \sum\limits_{k=1}^{n} J_k \left( \sum\limits_{i=0}^{n-k-1} s^j_i s^j_{i+k} \right)\right)$,
    \item $Y_{\eta_j, \eta_{j+1}} = - \sum\limits_{k=1}^n J_k \left( \sum\limits_{i=0}^{k-1} s^j_{n-i-1} s^{j+1}_{k-i-1} \right)$.
\end{itemize}

The terms in $X_{\eta_j}$ and $Y_{\eta_j, \eta_{j+1}}$ have the physical interpretations described below:

\begin{itemize}
    \item $- \dfrac{B}{2} \sum\limits_{i=0}^{n-1} s^{j}_i$ represents the energy contribution from the interactions between each spin in the block $\eta_j$ and the magnetic field $B$. For~$B > 0$, configurations tend to have all spins pointing up, while for $B < 0$ all spins pointing down are favored. Therefore, $B$ acts as a  type-$1$ periodicity parameter.
    \item $-\dfrac{1}{2} \sum\limits_{k=1}^{n} J_k \left( \sum\limits_{i=0}^{n-k-1} s^j_i s^j_{i+k} \right)$ represents the energy from the neighbor interactions between the spins within block $\eta_j$. For~$J_k > 0$, spins tend to align either all up or all down, favoring period-1 configurations. When $J_k < 0$, spin configurations of period-$2^R$ are prone to occur. Thus, $J_k$ serves as a type-2 periodicity parameter.
    \item $Y_{\eta_j, \eta_{j+1}}$ denotes the energy associated with interactions between spins in neighboring blocks $\eta_j$ and $\eta_{j+1}$. Since this term shares the same form and coupling as $-\dfrac{1}{2} \sum\limits_{k=1}^{n} J_k \left( \sum\limits_{i=0}^{n-k-1} s^j_i s^j_{i+k} \right)$, it leads to the same configuration patterns for corresponding values of $J_k$. Thus, $J_k$ again acts as a type-2 periodicity parameter.
\end{itemize}

\begin{figure}[H]
    
    \begin{minipage}{1.0\textwidth}
        
        \includegraphics[width=\linewidth]{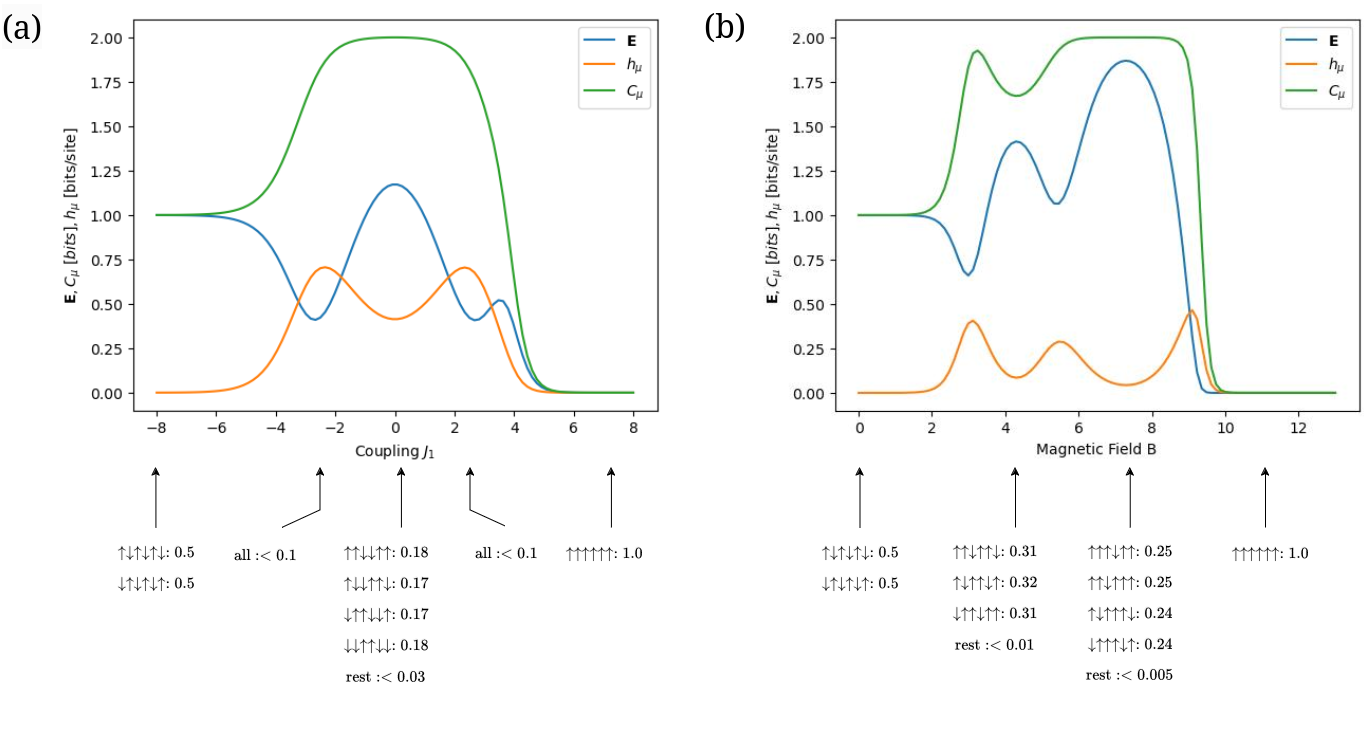}
    \end{minipage}
   
        \caption{(\textbf{a}) 
 $h_\mu$, \textbf{E} and $C_\mu$ vs. 
 $J_1$ for nnn Ising model with $J_2 = -1.2$, $B = 0.05$ and $T = 1$. (\textbf{b}) $h_\mu$, \textbf{E} and $C_\mu$ vs. $B$ for 3-range Ising model with $J_1 = -2.8$, $J_2 = -1.3$, $J_3 = -0.45$ and $T = 0.2$.}
        \label{fig:finite_range_ising_model_info_measures}
   
\end{figure}

The next step is to determine how effective information measures are at detecting and distinguishing configuration patterns  within typical configurations of finite-range Ising models. For~this, we start by considering a next-nearest neighbor Ising model with a moderately negative next-nearest-neighbor coupling $J_2=-1.2$, a very weak magnetic field $B=0.05$ and a low temperature $T=1$. Figure~\ref{fig:finite_range_ising_model_info_measures}a shows the model's information measures $h_{\mu}$, $\textbf{E}$ and $C_{\mu}$ as a function of the nearest-neighbor coupling $J_1 \in [-8, 8]$. To~assess the detection capability of these measures, typical configurations generated by the finite-range Boltzmann distribution at various values of $J_1$ are displayed below the horizontal~axis.

For a strongly negative nearest-neighbor coupling $J_1 \in [-8, -7)$, $h_{\mu}$ approaches zero, while  $\textbf{E} \approx 1$, together suggesting the presence of  period-$2$ typical configurations. In~this regime, the~ensemble exclusively adopts configurations that alternate between $\uparrow$ and $\downarrow$, confirming the period-$2$ pattern. These resulting configurations arise from the negative coupling $J_1$, which favors antiferromagnetic behavior~\cite{slotnick1951, zener1953}.

For a strongly positive nearest-neighbor coupling $J_1 \in [6, 8]$, all information measures approach zero, implying a period-$1$ typical configuration. The~resulting ``all-ups'' pattern observed at these values is consistent with these measures. This outcome is expected, as~the positive coupling $J_1$ drives the system toward ferromagnetic alignment~\cite{slotnick1951, zener1953}. 

For nearest-neighbor coupling $J_1=0.2$, the~system exhibits $h_{\mu}\approx 0.42$, $\textbf{E}\approx1.17$, and~reaches a maximum $C_{\mu}\approx2$. While $1 \leq \textbf{E} < 1.59$ would imply period-$3$ configurations in the absence of entropy rate, the~significant value of $h_{\mu}$ results in $C_{\mu} = 2$, pointing toward period-$4$ configurations. Consistently, at~these parameter values, we observe period-$4$ patterns in the typical configurations, including $\uparrow \uparrow \downarrow \downarrow$, $\uparrow \downarrow \downarrow \uparrow$, $\downarrow \uparrow \uparrow \downarrow$ and $\downarrow \downarrow \uparrow \uparrow$. Physically, this behavior can be understood as a result of the antiferromagnetic effect of $J_2$ being more dominant than the contributions from $B$ and $J_1$.

For $J_1=-2.5$ and $J_1=2.5$, all configurations have a probability of less than $0.1$. This indicates that, at~these parameter values, the~system does not have a typical configuration or preferred configuration pattern. Additionally, in~the regions $J_1 \in [-5, -1] \cup [2, 5]$, we observe that $C_{\mu}$ is not constant, but~exhibits significant variation. As~a result, these regions can be seen as \textit{configuration transition zones} where the typical configurations are shifting to new ones as the parameter of interest~varies. 

Now, consider a $3$-range Ising model with negative neighbor couplings of decreasing magnitude $J_1=-2.8$, $J_2=-1.3$, $J_3=-0.45$ and low temperature $T=0.2$. Figure~\ref{fig:finite_range_ising_model_info_measures}b shows the model's information measures $h_{\mu}$, $\textbf{E}$ and $C_{\mu}$ as a function of the magnetic field $B \in [0, 13]$. As~in
Figure~\ref{fig:finite_range_ising_model_info_measures}a, typical configurations at various values of $B$ are included below the horizontal~axis.

For a weak magnetic field $B \in [0, 0.75]$, we observe $h_{\mu} \approx 0$, $\textbf{E} \approx 1$, and~\mbox{$C_{\mu} \approx 1$}, indicating that only period-2 configurations are present, with~no possibility of other configurations, even as unlikely alternatives. This is further confirmed by the exclusivity of the alternating $\uparrow$ and $\downarrow$ configurations in this region. Moreover, for~a strong magnetic field $B \in [10, 13]$, all information measures approach zero, indicating that the system permits only period-1 configurations, consisting entirely of $\uparrow$ spins. This is further validated by the typical configurations calculated from the Boltzmann distribution. While the information measures and configuration patterns for these field ranges resemble those in Figure~\ref{fig:finite_range_ising_model_info_measures}a, they begin to differ in the intermediate range of $B$.

For a moderate magnetic field $B \approx 4.2$, we observe $h_{\mu} \approx 0.1$, $\textbf{E} \approx 1.4$, and~\mbox{$C_{\mu} \approx 1.7$}, indicating the presence of period-3 typical configurations. This is confirmed by the configurations calculated using the Boltzmann distribution. These results can be attributed to the competing effects between the antiferromagnetic couplings and the positive magnetic field~\cite{zarubin2019frustration, mutallib2022frustration, moessner2001frustration}. Moreover, in~Figure~\ref{fig:finite_range_ising_model_info_measures}a, the~probability of each non-typical configuration for $J_1 = 0.2$ is less than $0.03$, while in Figure~\ref{fig:finite_range_ising_model_info_measures}b, for~$B \approx 4.2$, the~probability of each non-typical configuration is less than $0.01$. The~lower value of $h_{\mu}$ for $B \approx 4.2$ in Figure~\ref{fig:finite_range_ising_model_info_measures}b, compared to that for $J_1 = 0.2$ in Figure~\ref{fig:finite_range_ising_model_info_measures}a, indicates that $h_{\mu}$ effectively captures the likelihood of non-typical~configurations.

For a strong magnetic field $B = 7.5$, the~typical configurations are period-4. Therefore, compared to Figure~\ref{fig:finite_range_ising_model_info_measures}a, Figure~\ref{fig:finite_range_ising_model_info_measures}b shows a greater variety of periodic patterns. Moreover, although~the 3-range model in Figure~\ref{fig:finite_range_ising_model_info_measures}b includes spins with two additional neighbors compared to the next-nearest neighbor model in Figure~\ref{fig:finite_range_ising_model_info_measures}a, it does not exhibit configuration patterns of periodicity higher than period-4. This captures how different parameters can limit or expand the diversity of configuration~patterns.

Notably, there is a dip around $B = 4.5$, where $B \approx |J_1 + J_2 + J_3|$. This suggests that, when the magnetic field and the coupling parameters are in a state of competing balance without a clear dominant effect, the~configuration patterns reach a complex yet not maximally intricate compromise. That is, their periodicity is higher than that of an antiferromagnet but still below the maximum possible within the range $B \in [0, 13]$.

\begin{figure}[H]

\begin{minipage}{0.9\textwidth} 

    \includegraphics[width=\linewidth]{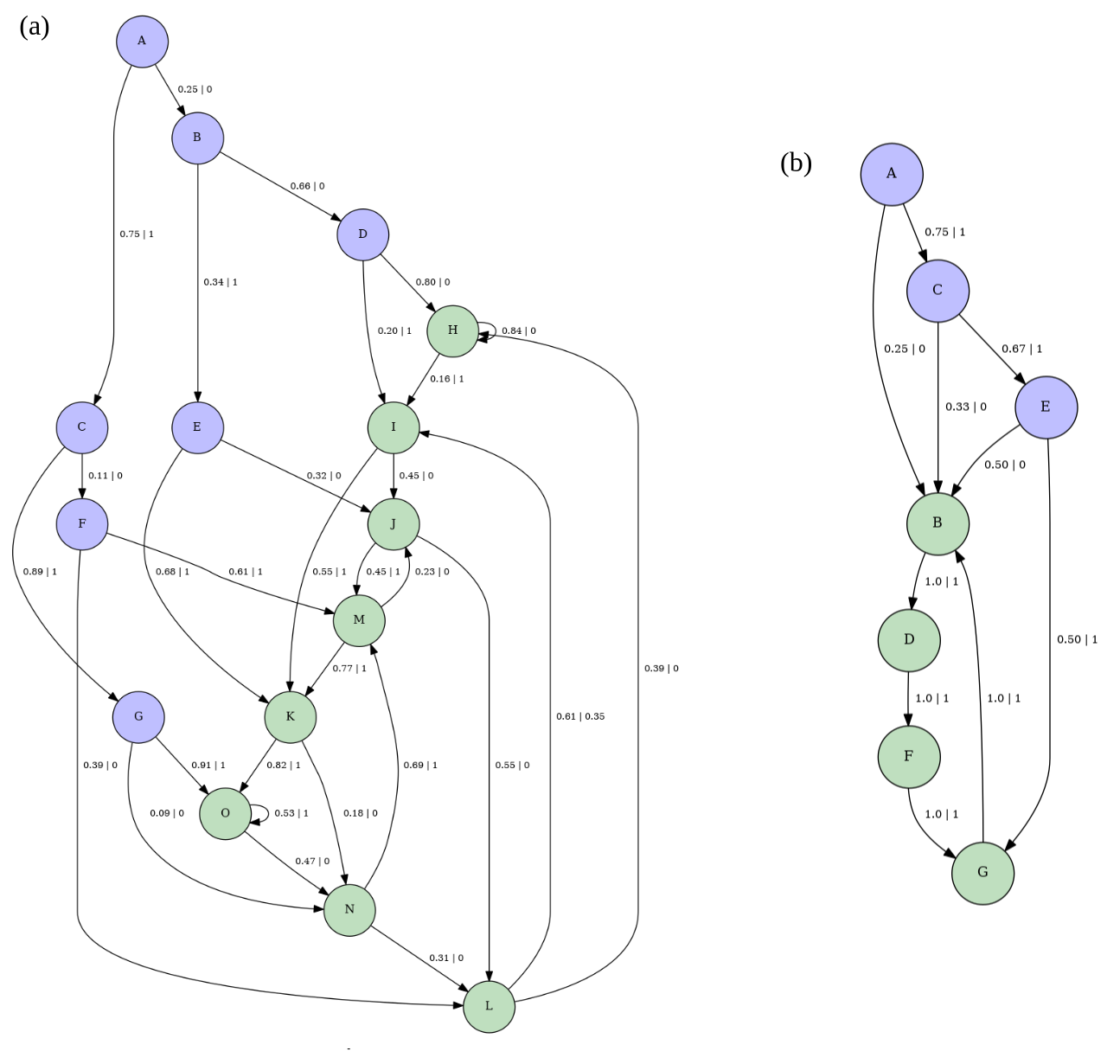}
\end{minipage}

    \caption{(\textbf{a}) 
 $\epsilon$-machine of 3-range Ising model with $B=0.2$, $T=4$, $J_1=1$, $J_2=1$ and $J_3=1$. (\textbf{b})~$\epsilon$-machine of 3-range Ising model with $B=8$, $T=0.2$, $J_1=-3$, $J_2=-2$ and $J_3=-2$.}\label{fig:finite_range_ising_model_epsilon_machines}

\end{figure}

Figure~\ref{fig:finite_range_ising_model_epsilon_machines} shows the $\epsilon$-machines for 3-range Ising models at fixed values of the coupling, temperature, and~magnetic field parameters.  In~panel (a), the~parameters are a weak magnetic field $B = 0.2$, a~moderate temperature $T = 4$, and~weak ferromagnetic couplings $J_1 = 1$, $J_2 = 1$, and~$J_3 = 1$. In~panel (b), the parameters are a strong magnetic field $B = 8$, a~low temperature $T = 0.2$, and~moderate antiferromagnetic couplings $J_1 = -3$, $J_2 = -2$ and $J_3 = -2$.

The $\epsilon$-machine in Figure~\ref{fig:finite_range_ising_model_epsilon_machines}a exhibits the maximum possible number of recurrent states, given by $2^R = 2^3 = 8$, where $R$ is the number of spins in a given spin block~\cite{feldman1998dnco}. Therefore, by~the definition of causal states, each spin block leads to a distinct future. This creates a one-to-one correspondence between spin blocks and causal states~\cite{feldman1998dnco}. Additionally, it has $7$ transient states, determined by $2^R - 1$ \cite{feldman1998dnco}. This indicates that $7$ spin variables must be observed before the next spin, allowing the observer to discern the precise typical configuration~pattern. 

Figure~\ref{fig:finite_range_ising_model_epsilon_machines}b shows fewer recurrent states compared to Figure~\ref{fig:finite_range_ising_model_epsilon_machines}a. This is due to the stronger magnetic field 
$B$ and lower temperature in Figure~\ref{fig:finite_range_ising_model_epsilon_machines}b, which bias typical configurations toward a period-1 pattern. Consequently, the~variety of possible typical spin configurations is reduced, limiting the range of possible futures. Moreover, Figure~\ref{fig:finite_range_ising_model_epsilon_machines}b exhibits only $3$ transient states. This can also be attributed to the bias toward period-1 configurations, as~fewer spins need to be observed to discern the typical configuration~pattern.

Furthermore, notice that Figure~\ref{fig:finite_range_ising_model_epsilon_machines}b has reduced connectivity compared to Figure~\ref{fig:finite_range_ising_model_epsilon_machines}a. Specifically, the~causal states in Figure~\ref{fig:finite_range_ising_model_epsilon_machines}a each have two outgoing transitions, while in Figure~\ref{fig:finite_range_ising_model_epsilon_machines}b, only transient states have two outgoing transitions, and~recurrent states have just one. This reduced connectivity is again a result of the low temperature, which limits the diversity of configuration patterns. Moreover, it can be further understood as a consequence of the balance between the magnetic field and coupling interactions, which leads to complex but not maximally intricate configuration~patterns. 

Ultimately, the~smaller size and reduced connectivity of the machine in Figure~\ref{fig:finite_range_ising_model_epsilon_machines}b, compared to Figure~\ref{fig:finite_range_ising_model_epsilon_machines}a, indicate that it performs less computation. Moreover, both panels in Figure~\ref{fig:finite_range_ising_model_epsilon_machines} illustrate that the number of causal states in a spin model does not always match the number of spin blocks; this occurs only when the model operates at maximum computational capacity. Instead, the~number of causal states varies based on internal factors like interaction couplings and external conditions such as the magnetic field and~temperature.

\subsection{Solid on Solid~Model}

In 1951, Burton, Frank, and~Cabrera (BFC) introduced a theory on the growth of real crystals in equilibrium, built upon earlier theories of perfect crystal growth~\cite{burton1951}. BFC posited that crystal growth is driven by the presence of steps on the crystal surface, with~the rate of growth determined by kinks in these~steps. 

In this context, a~step refers to the edge of an incomplete molecular layer on a crystal surface~\cite{burton1951}. The~interface between real crystals and their vapor is an example of a step~\cite{weeks1980}.  A~kink, on~the other hand, is an atomic site along a surface step where the atomic alignment at that point is~disrupted. 

In BFC’s theory, these kinks form on the surface at a specific temperature, referred to as the roughening temperature $T_R$. This prompted BFC to quantify surface roughness per molecule by comparing the potential energy per molecule at roughening and zero temperatures, as~shown in Equation~(\ref{eq:crystal_roughness}):
\begin{equation}
s = \dfrac{U_{R}-U_0}{U_0}
\label{eq:crystal_roughness}
\end{equation}

Here, $U_0$ and $U_{R}$ represent the potential energy per molecule at zero and roughening temperatures, respectively. The~difference $U_{R}-U_0$ is referred as the configurational potential energy, and~provided BFC with a gateway to model crystal surfaces as spin lattice~models.

They argued that for the $(001)$ surface of a simple cubic crystal, the~configurational potential energy is equivalent to the difference in potential energy between any two molecules~\cite{burton1951}. Consequently, this allows for the crystal surface to be modeled as a two-dimensional Ising model on a square lattice where each site is labeled by integer coordinates $x$ and $y$. Thus, the~potential energy between two molecules is given by the following:
\begin{equation}
u(\mu, \mu') = U|\mu - \mu'|
\label{eq:configurational_energy}.
\end{equation}

\begin{figure}[H]

\begin{minipage}{0.6\textwidth}
    
    \includegraphics[width=\linewidth]{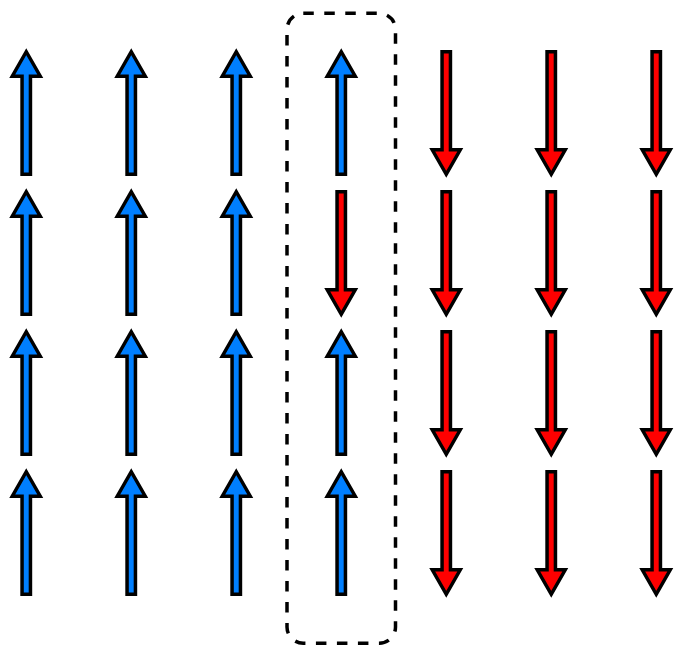}
\end{minipage}

    \caption{Illustration 
 of spin interactions in a 2D spin lattice with the leftmost and rightmost spins fixed to opposite values. The~dashed black lines highlight the induced 1D spin chain~interface.}
  \label{fig:sos_model}

\end{figure}

Moreover, by~focusing on kinks along the interface/step of a crystal with its vapor, the~problem can be simplified in two ways. First, all molecules on the surface to the left of the interface can be treated as spin up, and~those to the right as spin down~\cite{privman1988}. Second, these two regions can be regarded as forming a one-dimensional spin chain, reducing the Ising model from 2D to 1D~\cite{privman1988}, as~depicted in Figure~\ref{fig:sos_model}. This simplification is achieved by fixing the spins along the vertical boundaries at the extreme left $x=0$ to spin value $1$ and at the extreme right $x=x_{\text{high}}$ to spin value $-1$. These boundary conditions create a distinct transition in the lattice, where spin values switch from $1$ to $-1$. As~a result, the~Hamiltonian describing the configurational energy between two molecules is given by the following:
\begin{equation}
U|n_j - n_{j+1}|
\label{eq:spin_interaction_energy}
\end{equation}
where $n_j$ represents the number of leftmost up spins in row $j$ up to the interface at column~$i$.

If we further require that each occupied site sits directly above another occupied site—meaning no “overhangs” are allowed—then the one-dimensional spin chain meets the solid-on-solid condition~\cite{weeks1980}. 

Furthermore, an~attractive wall potential can be incorporated into the Hamiltonian of the configurational energy. Abraham demonstrated that this potential ``straightens'' the interface, provided that $x$ is restricted to lie in the right half of the plane, i.e.,~$0 \leq x \leq x_{\text{high}}$~\cite{abraham1979}. Following Privman~et~al.~\cite{privman1988}, a~simple attractive wall potential can be expressed~as:
\begin{equation}
W\delta_{1,n_y}
\label{eq:attractive_wall_potential}
\end{equation}

Moreover, an~additional external short-range potential can be included, represented~as follows:
\begin{equation}
E(n_y) \approx ce^{-an_y}, \quad a>0 
\label{eq:external_field}
\end{equation}

The resulting Hamiltonian for this system is given by Equation~\eqref{eq:sos_hamiltonian}:
\begin{equation}
E = \sum_y U|n_y - n_{y-1}| - W\delta_{1, n_y} + E(n_y)
\label{eq:sos_hamiltonian}
\end{equation}
where
\begin{itemize}
    \item $U |n_y - n_{y-1}|$ represents the energy cost of forming a kink in the interface. $U > 0$ biases the system toward period-1 configurations, while $U < 0$ favors alternating spins. Therefore, $U$ acts as a periodicity parameter of type 2.
    \item $- W\delta_{1, n_y}$ represents the energy associated with pinning the interface to the wall~\cite{privman1988}. For~$W > 0$, $n_y = 1$ prevails, while for $W < 0$, $n_y = 0$ dominates. In~both cases, the~system favors period-1 configurations. Thus, $W$ serves as a type 1 periodicity parameter.
    \item $E(n_y)$ represents the energy contribution from an external field that influences the interface's orientation or tilt~\cite{privman1988}. For~$E > 0$, an~interface made up of $1$s is favored, while for $E < 0$, an~interface made up of $0$s is preferred. Therefore, the~parameters in this term function as type-1 periodicity parameters.
\end{itemize}

In what follows, we restrict $n_y \in \{0, 1\}$, so that under 
$s_y = 2n_y - 1$ the SOS Hamiltonian is equivalent to a nearest-neighbor 1D Ising chain. We compute the probabilities needed for the information measures and 
$\epsilon$-machines directly from the Boltzmann distribution associated with Equation~\eqref{eq:sos_hamiltonian} using the transfer-matrix~method. 

\begin{figure}[H]

\begin{minipage}{0.95\textwidth}
    
    \includegraphics[width=\linewidth]{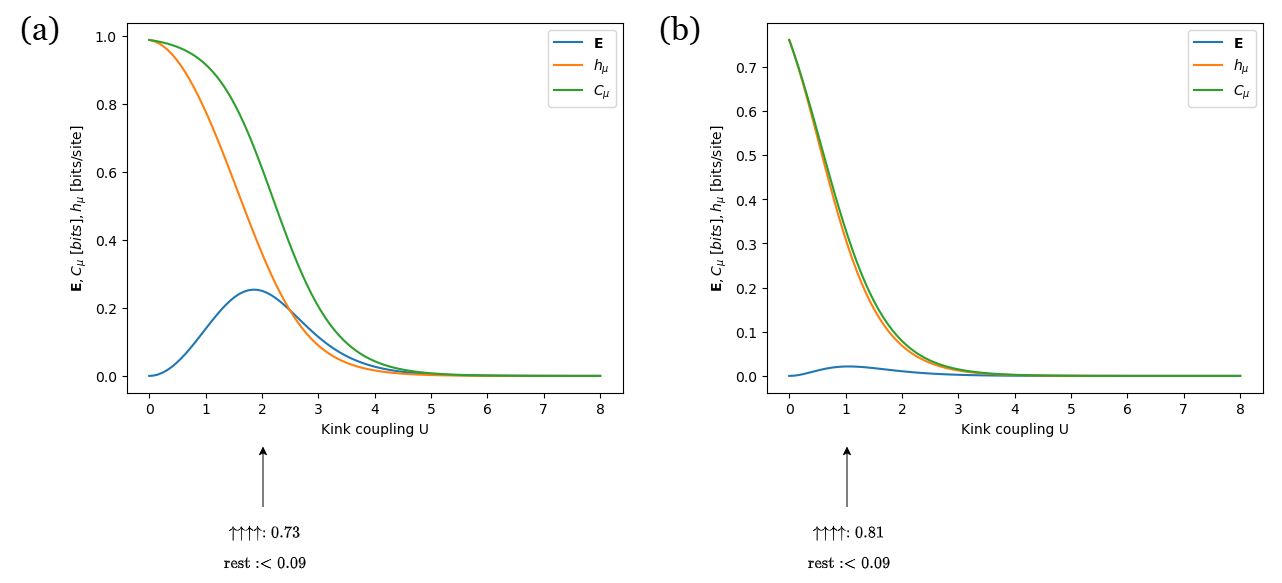}
\end{minipage}

    \caption{(\textbf{a}) 
    $h_{\mu}$, $\textbf{E}$ and $C_{\mu}$ vs. $U$ for SOS model with $W=0$, $V=e^{-n_y}$ and $T=1$. (\textbf{b})~$h_{\mu}$, $\textbf{E}$ and $C_{\mu}$ vs. $U$ for SOS model with $W=1$, $V=e^{-n_y}$ and $T=1$.}\label{fig:sos_model_info_measures}

\end{figure}

We now aim to compare how turning the pinning wall $W$ on and off affects both the configurations and information measures of the SOS model. For~this comparison, we consider an SOS model with low temperature $T=1$, external potential $V=e^{-n_y}$ and pinning wall potential $W=0 \ \text{or} \ 1$. Figure~\ref{fig:sos_model_info_measures}  displays the information measures of the SOS model as the kink coupling $U$ varies. In~Figure~\ref{fig:sos_model_info_measures}a,b, the~pinning wall $W$ is set to 0 and 1, respectively.

In both panels of Figure~\ref{fig:sos_model_info_measures}, $C_{\mu}$ ranges from $0$ to $1$, indicating typical configurations of either period-1 or period-2.  In~both figures, even a slight increase in the kink coupling above zero causes $C_{\mu}$ to reach its peak value. This behavior aligns with the Gibbsean assumption that a low cost of forming kinks makes non-uniform configurations—that is, non-period-1 configurations—more likely to occur~\cite{burton1951}. 

In Figure~\ref{fig:sos_model_info_measures}a, $C_{\mu}$ reaches its peak just below 1, whereas in Figure~\ref{fig:sos_model_info_measures}b, it peaks around 0.75. Moreover, for~$0 < U < 5$, $C_{\mu}$ stays higher in Figure~\ref{fig:sos_model_info_measures}a than in Figure~\ref{fig:sos_model_info_measures}b. This sustained higher value of $C_{\mu}$ in Figure~\ref{fig:sos_model_info_measures}a compared to Figure~\ref{fig:sos_model_info_measures}b is in line with the SOS Hamiltonian, which suggests that biasing the interface toward the pinning wall increases the likelihood of the interface becoming flat, that is, period-1~\cite{burton1951}.

In Figure~\ref{fig:sos_model_info_measures}a, $\textbf{E}$ peaks around $\textbf{E}=0.26$ at $U=1.8$, while in Figure~\ref{fig:sos_model_info_measures}b, it peaks around $\textbf{E}=0.04$ at $U=1$. This suggests that more spins need to be observed to determine the configuration pattern of the SOS model in Figure~\ref{fig:sos_model_info_measures}a compared to Figure~\ref{fig:sos_model_info_measures}b. This is consistent with period-1 configurations being more likely in Figure~\ref{fig:sos_model_info_measures}b, as~these configurations do not require observing any~spins.

At the $\textbf{E}$ peak in Figure~\ref{fig:sos_model_info_measures}a, $C_{\mu} = 0.75$, while at that of Figure~\ref{fig:sos_model_info_measures}b, $C_{\mu} = 0.35$. This implies that period-2 configurations are more likely to occur in Figure~\ref{fig:sos_model_info_measures}a compared to Figure~\ref{fig:sos_model_info_measures}b. This aligns with typical period-1 configurations being less prevalent and, conversely, non-typical period-2 configurations being more frequent in Figure~\ref{fig:sos_model_info_measures}a compared to Figure~\ref{fig:sos_model_info_measures}b. Moreover, this is consistent with the physical expectation that biasing the interface to be attracted to the wall increases the likelihood that it becomes flat, thereby raising the probability of a period-1~configuration.

Furthermore, in~both panels of  Figure~\ref{fig:sos_model_info_measures}, as~the kink coupling $U$ increases, $h_{\mu}$ decreases. This trend is expected, as~the higher cost of kink formation makes non-period-1 configurations less likely, thereby reducing the uncertainty of the next observed spin. The~decrease occurs more rapidly in Figure~\ref{fig:sos_model_info_measures}b compared to Figure~\ref{fig:sos_model_info_measures}a. This can be explained by the presence of the pinning wall, which further encourages the dominance of flat, period-1~configurations.

\begin{figure}[H]

\begin{minipage}{0.9\textwidth}
    
    \includegraphics[width=\linewidth]{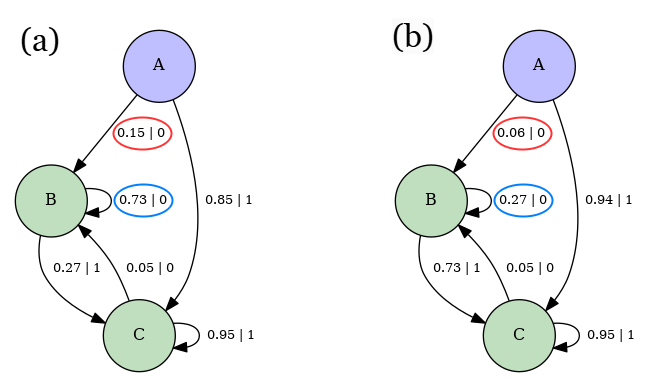}
\end{minipage}

    \caption{(\textbf{a}) 
    $\epsilon$-machine for SOS model with $U=2$, $W=0$, $V=e^{-n_y}$, $T=1$ and $C_{\mu} \approx 0.61$. (\textbf{b})~$\epsilon$-machine for SOS model with $U=1$, $W=1$, $V=e^{-n_y}$, $T=1$ and $C_{\mu} \approx 0.33$.}\label{fig:sos_model_epsilon_machines}

\end{figure}

Figure~\ref{fig:sos_model_epsilon_machines}a,b show the $\epsilon$-machines corresponding to the $\textbf{E}$ peaks of  Figure~\ref{fig:sos_model_info_measures}a,b. Both $\epsilon$-machines feature two recurrent states and one transient state. However, as~circled in red, the~probability of transitioning from state $A$ to state $B$ while outputting symbol $0$ in Figure~\ref{fig:sos_model_epsilon_machines}a is more than twice as high as in Figure~\ref{fig:sos_model_epsilon_machines}b. Moreover, as~circled in blue, the~probability of transitioning from state $B$ to state $B$ while outputting symbol $0$ decreases from $0.73$ in
Figure~\ref{fig:sos_model_epsilon_machines}a to $0.27$ in Figure~\ref{fig:sos_model_epsilon_machines}b. This bias towards period-1 configurations of the machine in Figure~\ref{fig:sos_model_epsilon_machines}b suggests that it is easier for the machine in Figure~\ref{fig:sos_model_epsilon_machines}b to synchronize than the one in in Figure~\ref{fig:sos_model_epsilon_machines}a, which aligns with the fact that the $\textbf{E}$ is higher for the machine in Figure~9b while both machines have similar values of $h_{\mu}$ (approximately 0.36 for Figure~\ref{fig:sos_model_epsilon_machines}a and 0.31 for Figure~\ref{fig:sos_model_epsilon_machines}b). Moreover, the~outgoing transition probabilities from the transient state in Figure~\ref{fig:sos_model_epsilon_machines}b are less uniform than those in Figure~\ref{fig:sos_model_epsilon_machines}a. This suggests that while both machines can identify the ``all-ups'' configuration without observing any spins, this configuration is more representative of the machine in Figure~\ref{fig:sos_model_epsilon_machines}b than of the one in Figure~\ref{fig:sos_model_epsilon_machines}a. Computationally, this means that the behavior of the machine in Figure~\ref{fig:sos_model_epsilon_machines}b more closely resembles that of a single-state machine that exclusively outputs symbol $1$.

Lastly, note that the machines for the SOS model in Figure~\ref{fig:sos_model_info_measures} and the nearest-neighbor Ising model in the Appendix \ref{appendix:nn_ising_machine} share similar recurrent states, transient states, and~connectivity, but~have different state transition probabilities. This suggests that the $\epsilon$-machines offer a constructive framework for comparing the structures of different spin models and examining their similarities and~differences.

\subsection{Three-Body~Model}

Thermal desorption is the process of heating a solid surface to release a portion of its molecules~\cite{wang2012desorption, aparicio2022desorption}. 
The defining characteristic of this process is its kinetics, which are described by the desorption rate and the desorption rate constant, as~outlined in~\cite{zhdanov1981lattice} and presented in Equations~(\ref{eq:desorption_rate}) and (\ref{eq:desorption_rate_constant}), respectively. These two key equations are directly connected to experiment, as~at sufficiently high pumping rates, the~desorption rate equals the desorbant's pressure~\cite{redhead1962}:
\begin{equation}
\dfrac{d \theta}{dt} = -k_d \theta
\label{eq:desorption_rate}
\end{equation}
\begin{equation}
k_d = \nu \sum\limits_i  P_{A,i} \exp{\left(-\frac{E_d(0)- E_i}{T} \right)}
\label{eq:desorption_rate_constant}
\end{equation}

Detecting the temperatures at which desorption is greatest and identifying the qualitative properties of desorption at these values is crucial for various applications~\cite{redhead1962}. To~achieve these objectives, the~negative desorption rate is plotted against temperature to obtain the ``desorption spectrum''~\cite{redhead1962}. The~peaks in this spectrum indicate the temperatures at which the desorption rate is highest. These peaks vary in width, height, and~location depending on the coverage, temperature, and~material examined. For~the case of the desorption spectrum of CO from Ni, Pd, Pt, Rh, and Ru closed-packed faces of single crystals, two distinguishing qualitative features arise, as~demonstrated by Morris et al.~\cite{morris1984}:
\begin{enumerate}
    \item The splitting of thermal desorption peaks becomes progressively weaker as one goes from Ni to Ru.
    \item The integral intensities of the peaks are distinct.
\end{enumerate}

\begin{figure}[H]

\begin{minipage}{0.8\textwidth}
    
    \includegraphics[width=\linewidth]{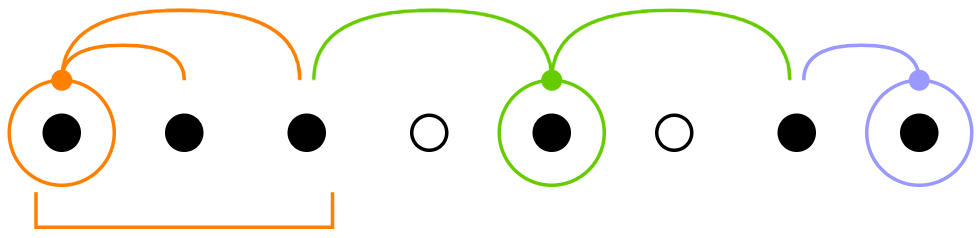}
\end{minipage}

    \caption{Illustration of spin interactions in three-body models: nearest-neighbor (purple), next-nearest neighbor (green), and~three-body (orange) couplings.}\label{fig:three?body_model}\label{fig:three_body_model}

\end{figure}

While nearest-neighbor (nn) and next-nearest-neighbor (nnn) spin models had been used to model thermal desorption~\cite{zhdanov1986}, they did not capture the aforementioned properties. Myshlyavtsev~et~al. addressed this limitation by incorporating a three-body term in the spin Hamiltonian, which effectively models these characteristics~\cite{myshlyavtsev1989}. The~resulting three-body model removes the assumption of paired interactions~\cite{zhdanov1986}, providing a more accurate account of the CO desorption process from metal surfaces. The~1D model is exactly solvable and, if~lateral interactions are anisotropic, sufficient to capture thermal desorption, making it of theoretical and practical interest, respectively~\cite{myshlyavtsev1989}. The~spin interactions in the 1D three-body model are illustrated in Figure~\ref{fig:three_body_model}. The~Hamiltonian for this model is given in Equation~\eqref{eq:three_body_hamiltonian}, and~the corresponding transfer matrix is detailed in Appendix \ref{appendix:three_body_transfer_matrix}.
\begin{equation}
E(s_i, s_{i+1}, s_{i+2}) = \sum_i - J_1 s_i s_{i+1} - J_2 s_i s_{i+2} - J_{\text{tb}} s_i s_{i+1} s_{i+2} 
\label{eq:three_body_hamiltonian}
\end{equation}
where
\begin{itemize}
    \item $-J_1 s_i s_{i+1}$ is the term associated with the nearest-neighbor coupling. For~$J_1>0$, the~model induces period-1 configurations, while for $J_1<0$, the~model induces period-2 configurations. Thus, $J_1$ serves as a type-2 periodicity parameter.
    \item $-J_1 s_i s_{i+2}$ is the energy contribution of the next-nearest-neighbour coupling. When $J_2 > 0$, the~model tends toward period-1 configurations, whereas for $J_2 < 0$, it leans toward period-4 configurations. Therefore, $J_2$ acts as a periodicity parameter of type $2$.
    \item $-J_{\text{tb}} s_i s_{i+1} s_{i+2}$ is the expression that represents the three-body interaction. When $J_{\textit{tb}} > 0$, the~configurations are biased toward a period-1 pattern, while $J_{\textit{tb}} < 0$ favors period-$4$ configurations. As~a result, $J_{\text{tb}}$ functions as a type 2 periodicity parameter. 
\end{itemize}

\begin{figure}[H]

\begin{minipage}{0.95\textwidth}
    
    \includegraphics[width=\linewidth]{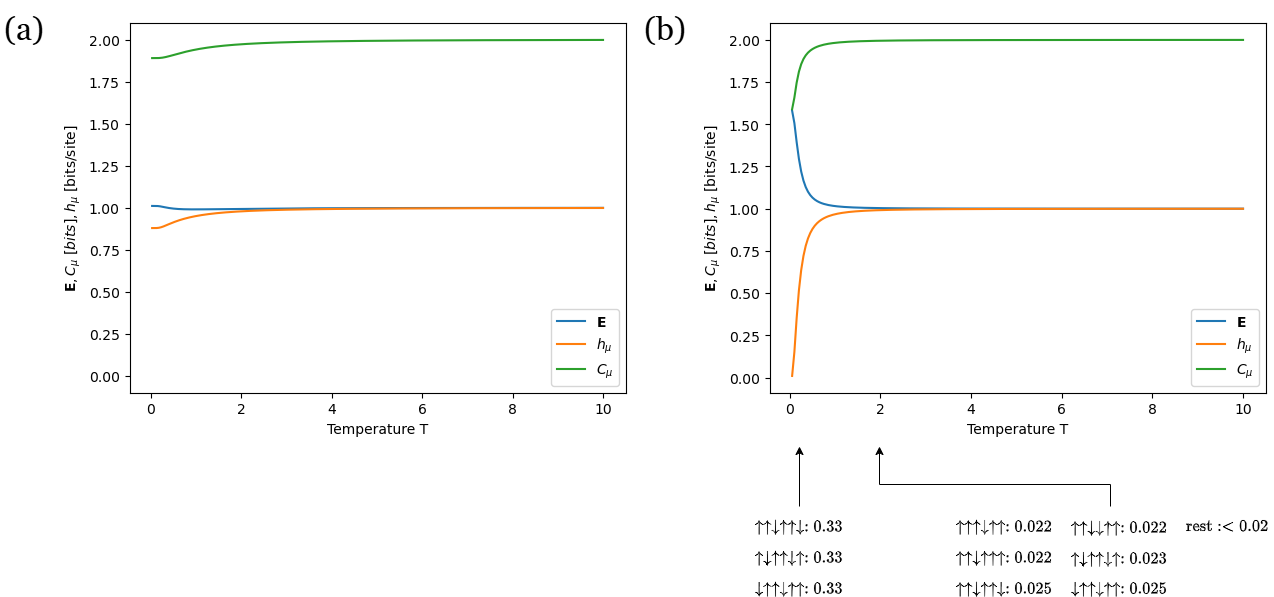}
\end{minipage}
    \caption{(\textbf{a}) 
    $h_{\mu}$, $\textbf{E}$ and $C_{\mu}$ vs. $T$ for three body model with $J_1=0$, $J_2=0$ and $J_t=-1$. (\textbf{b})~$h_{\mu}$, $\textbf{E}$ and $C_{\mu}$ vs. $T$ for three body model with $J_1=1$, $J_2=0$ and $J_t=-1$.}\label{fig:three_body_model_info_measures}

\end{figure}

The purpose of Figure~\ref{fig:three_body_model_info_measures} is to illustrate how turning the nearest-neighbor coupling on and off in a three-body model affects both its configurations and information measures as a parameter of interest varies. Temperature is chosen as that parameter because it plays a key role in thermal desorption applications, where the goal is to identify the temperature that maximizes desorption~\cite{redhead1962, myshlyavtsev1989}. In~both panels, the~next-nearest-neighbor coupling $J_2$ is set to $0$ to highlight the role of the nearest-neighbor coupling $J_1$, while the three-body coupling $J_{\text{tb}}$ is set to $-1$. However, in~Figure~\ref{fig:three_body_model_info_measures}a, the~nearest-neighbor coupling $J_1$ is set to $0$, whereas in Figure~\ref{fig:three_body_model_info_measures}b, it is set to $1$.

In both Figure
~\ref{fig:three_body_model_info_measures}a and Figure~\ref{fig:three_body_model_info_measures}b, $C_{\mu}$ increases and reaches its maximum value of $C_{\mu} = 2$ as the temperature $T$ rises, but~the starting values differ. In~Figure~\ref{fig:three_body_model_info_measures}a, $C_{\mu}$ begins around $1.9$, whereas in Figure~\ref{fig:three_body_model_info_measures}b, it starts at approximately $C_{\mu} \approx 1.58$. This suggests that at low temperature values, the~typical configurations in Figure~\ref{fig:three_body_model_info_measures}a are period-4, and~in Figure~\ref{fig:three_body_model_info_measures}b, they are period-3. This difference can be attributed to the fact that Figure~\ref{fig:three_body_model_info_measures}b involves competing couplings, whereas Figure~\ref{fig:three_body_model_info_measures}a does not, as~it only includes the three-body coupling. In~particular, in~both Figure
~\ref{fig:three_body_model_info_measures}a and Figure~\ref{fig:three_body_model_info_measures}b, the~three-body coupling $J_{\text{tb}}$ biases configurations toward a period-4 pattern. However, in~Figure~\ref{fig:three_body_model_info_measures}b, the~ferromagnetic coupling $J_1$ also biases configurations toward a period-1 pattern. The~competition leads to a compromise, resulting in period-3 configurations. This is consistent with the low-temperature typical configurations calculated using the Boltzmann distribution, which are shown below the horizontal axis in Figure~\ref{fig:three_body_model_info_measures}b.

Moreover, the~nearest-neighbor coupling significantly reduces the uncertainty in predicting the next spin by expanding the neighborhood of spins that each state affects. This leads to a lower 
$h_{\mu}$ at very low temperatures in Figure~\ref{fig:three_body_model_info_measures}b compared to Figure~\ref{fig:three_body_model_info_measures}a. This prevents $C_{\mu}$ in Figure~\ref{fig:three_body_model_info_measures}b from being strongly influenced by $h_{\mu}$ at very low~temperatures. 

Furthermore, although~$C_{\mu}$ is higher in Figure~\ref{fig:three_body_model_info_measures}a than in Figure~\ref{fig:three_body_model_info_measures}b at low temperatures, $\textbf{E}$ is lower in Figure~\ref{fig:three_body_model_info_measures}a compared to Figure~\ref{fig:three_body_model_info_measures}b at the same temperatures. This implies that while typical configurations in Figure~\ref{fig:three_body_model_info_measures}a at very low temperatures exhibit greater periodicity than those in Figure~\ref{fig:three_body_model_info_measures}b (period-4 versus period-3), the~observer must examine more spin variables to discern the configuration pattern in Figure~\ref{fig:three_body_model_info_measures}b. While this might seem to suggest that patterns in Figure~\ref{fig:three_body_model_info_measures}b are easier to discern than those in Figure~\ref{fig:three_body_model_info_measures}a, the~uncertainty per spin in Figure~\ref{fig:three_body_model_info_measures}b is significantly higher. Specifically, $h_\mu \approx 0$ for Figure~\ref{fig:three_body_model_info_measures}a, whereas $h_\mu \approx 0.9$ for Figure~\ref{fig:three_body_model_info_measures}b. This substantial difference makes an information-theoretic approach based on excess entropy $\mathbf{E}$ insufficient for determining the ease of synchronization. We will soon address this by examining the computational properties of the three-body~models.

Notably, the~information measures of the three-body model reveal new features that were absent in previously studied spin models. For~instance, unlike the dependence of $\textbf{E}$ on temperature in the nearest-neighbor Ising model, where $\textbf{E}$ decays to $0$ as $T$ increases (as shown in Ref.~\cite{feldman1998dnco} and Appendix~\ref{appendix:nn_ising_info_vs_temp}), $\textbf{E}$ for the three-body model remains nonzero even at high temperatures. Moreover, even though there is no magnetic field $B$ in \mbox{Figure~b,} 
 the~information measures are not flat across the temperature range. This suggests that a diversity of configuration patterns is possible whenever competing parameters are present, regardless of their specific nature, which further reinforces the usefulness of our classification of parameter types. Ultimately, the~information measures plots in \mbox{Figures
~\ref{fig:finite_range_ising_model_info_measures},~\ref{fig:sos_model_info_measures} and~\ref{fig:three_body_model_info_measures}} suggest that different spin models give rise to distinct configuration patterns and structural~behavior.

\begin{figure}[H]

\begin{minipage}{0.9\textwidth}
    
    \includegraphics[width=\linewidth]{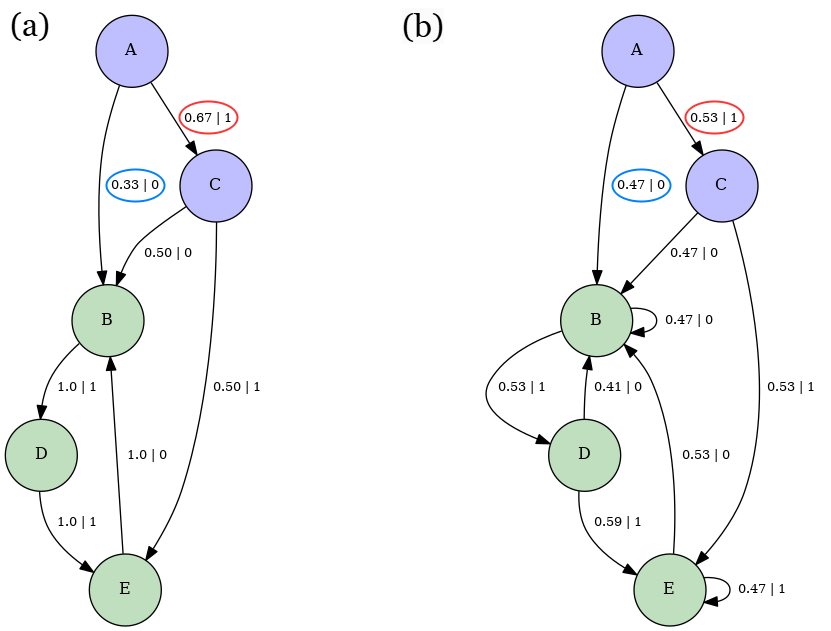}
\end{minipage}

    \caption{(\textbf{a}) 
    $\epsilon$-machine for three body model with $J_1=-1$, $J_2=0$, $J_t=1$, $T=0.025$. (\textbf{b}) $\epsilon$-machine for three body model with $J_1=-1$, $J_2=0$, $J_t=1$, $T=2$}\label{fig:three_body_model_epsilon_machines}

\end{figure}

Figure~\ref{fig:three_body_model_epsilon_machines} aims to illustrate the structural changes in the $\epsilon$-machine of a three-body model with competing couplings as the temperature increases. The~plots in Figure~\ref{fig:three_body_model_epsilon_machines}a and Figure~\ref{fig:three_body_model_epsilon_machines}b depict the $\epsilon$-machines corresponding to Figure~\ref{fig:three_body_model_info_measures}b at a very low temperature $T=0.025$ and a low temperature $T=2$, respectively.

The outgoing probabilities from the causal transient state $A$ to the transient states $B$ and $C$ in Figure~\ref{fig:three_body_model_epsilon_machines}a, which are circled in red and blue, are less uniform than those in Figure~\ref{fig:three_body_model_epsilon_machines}b. This implies that the $\epsilon$-machine in Figure~\ref{fig:three_body_model_epsilon_machines}a is easier to synchronize than the one in Figure~\ref{fig:three_body_model_epsilon_machines}b. At~first, this may seem inconsistent with their excess entropy values, given that $\textbf{E}\approx1.58$ for Figure~\ref{fig:three_body_model_epsilon_machines}a and $\textbf{E}=1$ for Figure~\ref{fig:three_body_model_epsilon_machines}b, as~shown in Figure~\ref{fig:three_body_model_info_measures}b. However, this apparent contradiction is resolved by observing the significantly higher value of $h_\mu$ in Figure~\ref{fig:three_body_model_epsilon_machines}b compared to Figure~\ref{fig:three_body_model_epsilon_machines}a, where $h_\mu \approx 1$ for Figure~\ref{fig:three_body_model_epsilon_machines}b and $h_\mu \approx 0$ for Figure~\ref{fig:three_body_model_epsilon_machines}a. As~a result, while discerning configuration patterns in Figure~\ref{fig:three_body_model_epsilon_machines}a may require an additional spin, the~much higher uncertainty in predicting the next spin in Figure~\ref{fig:three_body_model_epsilon_machines}b outweighs this requirement, making synchronization more challenging in Figure~\ref{fig:three_body_model_epsilon_machines}b than in Figure~\ref{fig:three_body_model_epsilon_machines}a. This uncertainty is further supported by the fact that typical configurations for Figure~\ref{fig:three_body_model_epsilon_machines}b are much less probable than those in Figure~\ref{fig:three_body_model_epsilon_machines}a. Specifically, the~highest probability for a typical configuration in Figure~\ref{fig:three_body_model_epsilon_machines}a is $0.33$, whereas in Figure~\ref{fig:three_body_model_epsilon_machines}b, it is only $0.025$. This contrast highlights how the computational approach provided by $\epsilon$-machines offers a more nuanced perspective on synchronization than the randomness-agnostic viewpoint of excess entropy $\textbf{E}$.

Moreover, the~recurrent part of Figure~\ref{fig:three_body_model_epsilon_machines}a is much less connected than that of \mbox{Figure~\ref{fig:three_body_model_epsilon_machines}b.}  In~the $\epsilon$-machine for Figure~\ref{fig:three_body_model_epsilon_machines}a, each recurrent causal state has only one outgoing transition with probability $1.0$. In~contrast, the~recurrent states in Figure~\ref{fig:three_body_model_epsilon_machines}b each have two outgoing transitions, both with probabilities close to $0.50$.
Furthermore, Figure~\ref{fig:three_body_model_epsilon_machines}b includes self-loops that enable it to recognize period-1 configurations consisting entirely of $0$s or $1$s, a~feature absent in Figure~\ref{fig:three_body_model_epsilon_machines}a. This indicates that the machine in \mbox{Figure~\ref{fig:three_body_model_epsilon_machines}b} generates a greater variety of spin configurations compared to the one in Figure~\ref{fig:three_body_model_epsilon_machines}a. This observation is consistent with the fact that at $T=0.025$, there are only three typical configurations, whereas at $T=2$, there are~six.

Lastly, the~number of recurrent causal states, along with the low connectivity of the machine in Figure~\ref{fig:three_body_model_epsilon_machines}a, suggests that it can support configurations with periods of up to~$3$. In~contrast, the~machine in Figure~\ref{fig:three_body_model_epsilon_machines}b, which has the same number of recurrent causal states but higher connectivity, permits configurations with periods of up to $4$. The~typical configurations in Figure~\ref{fig:three_body_model_info_measures}b reflect this pattern, as~Figure~\ref{fig:three_body_model_epsilon_machines}b accommodates both period-4 and period-3 configurations, whereas Figure~\ref{fig:three_body_model_epsilon_machines}a only supports period-3 configurations.  Ultimately, this comparison of $\epsilon$-machines underscores the importance of considering not only typical configurations but also their probabilities when developing a computation-theoretic account of spin~patterns.

\section{Conclusions}

What, then, is a pattern in statistical mechanics? If one recasts the mechanism generating a system's structure as an information processor, the~answer for the one-dimensional spin models studied here is clear: the $\epsilon$-machine. To~support this perspective, we began by introducing computational mechanics and its application to statistical mechanics in a conceptual manner with only the necessary amount of mathematics. We then defined typical configurations and typical configuration patterns as the most likely configurations and configuration patterns within an ensemble. Furthermore, we classified the parameters of spin models according to the type of behavior they give rise to. 

Using this framework, we computed typical configurations from the embedded Boltzmann distribution and compared them to those implied by information measures and $\epsilon$-machines for three different spin models: the finite-range Ising model, the~SOS model, and~the three-body model. Our findings confirmed consistency between the results, establishing the $\epsilon$-machine as a representation of the Boltzmann distribution's ensemble patterns. Moreover, our analysis showed that information measures and $\epsilon$-machines offer a detailed and nuanced characterization of typical configuration patterns, allowing us to distinguish between them and identify their shared~features.

In the finite-range Ising model, the~information plots show that $C_{\mu}$ serves as a simple visual indicator of regions where no typical configurations exist. These regions, distinguished by the non-flat behavior of $C_{\mu}$, are what we refer to as transition zones. Furthermore, $C_{\mu}$ captures the fact that different parameters influence the diversity of configuration patterns, and~consequently, the~computational demands. For~instance, a~dominant antiferromagnetic $J_2$ coupling maximizes computation, while competing effects between $B$ and antiferromagnetic $J_1$ lead to high but constrained~computation.

Moreover, the~$\epsilon$-machines of the finite-range Ising model provide a more refined perspective on the computational differences arising from varying parameters. For~instance, the~high but constrained computation observed in the three-range Ising model with a high magnetic field and low temperature is represented by fewer causal states and lower connectivity compared to a system with a low magnetic field and moderate temperature. This distinction provides a more nuanced understanding of what it means for a system to require more or less computation. Additionally, the~analysis shows that the number of causal states cannot simply be inferred from properties such as the number of neighbors a given spin has or the magnitude and sign of the~parameters. 

In the SOS model, $C_\mu$ allows us to quantify the reduction in computational effort caused by turning the wall on, even when the typical configuration remains unchanged. Furthermore, the~observation that maximum $C_{\mu}$ occurs at very low kink coupling demonstrates that the peak of maximum $C_{\mu}$ varies depending on the specific parameter under consideration. Moreover, the~$\epsilon$-machines of the SOS model show that turning on the wall parameter reduces the uniformity of the outgoing transition probabilities from the start state. This indicates that the typical configuration becomes more likely as the wall parameter becomes nonzero. In~computational terms, when the wall is fully activated, the~machine becomes more similar to a single-state machine that solely outputs $1$. More broadly, the~machines from this case study, along with those of the nearest-neighbor Ising model, demonstrate that $\epsilon$-machines provide a unified framework for identifying computational similarities (such as the number of states and connectivity) and differences (such as transition probabilities) between two distinct spin~models.

The information measures of the three-body models, both with and without nearest-neighbor coupling, are not monotonically dependent. Specifically, a~high $\textbf{E}$ or $h_{\mu}$ is shown to not necessarily imply a high or low $C_{\mu}$. The~information plots of these three body models, along with those of the finite-range Ising models, capture how different spin models produce distinct configuration patterns when the same parameter, in~this case temperature, is varied. Furthermore, the~$\epsilon$-machines of the three-body model with nearest-neighbor coupling provide an effective framework for identifying computational similarities and differences in the spin model as a parameter, such as temperature, changes. As~temperature increases, typical configurations become more periodic, but~also less likely to occur. The~$\epsilon$-machines capture this behavior by making the outgoing probabilities from transient causal states more uniform while increasing connectivity in the recurrent portion. This suggests that as typical configuration patterns become more periodic and less likely, they also become harder to discern overall. Notably, this result highlights the limitations of an information-theoretic perspective on synchronization—while useful, it remains incomplete without a computational viewpoint. This insight sheds light on subtle structural differences between systems that, despite having the same number of recurrent and transient causal states, exhibit distinct dynamical~behaviors.

Ultimately, information theory and computational mechanics provide powerful tools for defining patterns in Boltzmann ensembles and for comprehensively characterizing the typical configurations generated by the Boltzmann distribution. They also enable a unified way of examining similarities and differences in the structure and patterns of a spin model under varying parameters and across different spin models. This perspective connects the abstract formalism of information theory and automata theory with the concrete physical models of statistical mechanics, providing a constructive and effective language to describe patterns in statistical~mechanics. 

\vspace{6pt}

\funding{This research was funded by the Foundational Questions Institute and by the Faggin Presidential
Chair~Fund.}

\dataavailability{The code supporting this study is available at: 
 \url{https://github.com/omalagui/spin_patterns} (accessed on 13 January 2026).} 

\acknowledgments{The author is grateful to Josh Deutsch, Jim Crutchfield, Anthony Aguirre, Zara Brandt, Evan Frangipane, Vidyesh Rao Anisetti and Jordan Scharnhorst for helpful feedback and insightful~conversations. }

\conflictsofinterest{The author declares no conflicts of~interest.} 

\appendixtitles{yes} 
\appendixstart
\appendix

\section{Concept of ``State'' in Theory of Computation and Its Formalization in Computational~Mechanics} \label{appendix:TOC_concept_of_state}

In automata theory, abstract machines—the primary objects of study—are formalized in terms of ``states.'' However, the~concept of ``state'' itself lacks an explicit mathematical definition. Even at a conceptual level, a~``state'' is rarely defined. One notable exception appears in~\cite{hopcroft2001} (pp. 2--3), where a state is defined as the relevant portion of a system's history.  Although~the purpose of this relevance is not specified, it is illustrated through the example of a very simple finite-state machine: an on-off switch, shown~below.

\begin{figure}[H]
    
    \begin{minipage}{0.4\textwidth}
        
        \includegraphics[width=\linewidth]{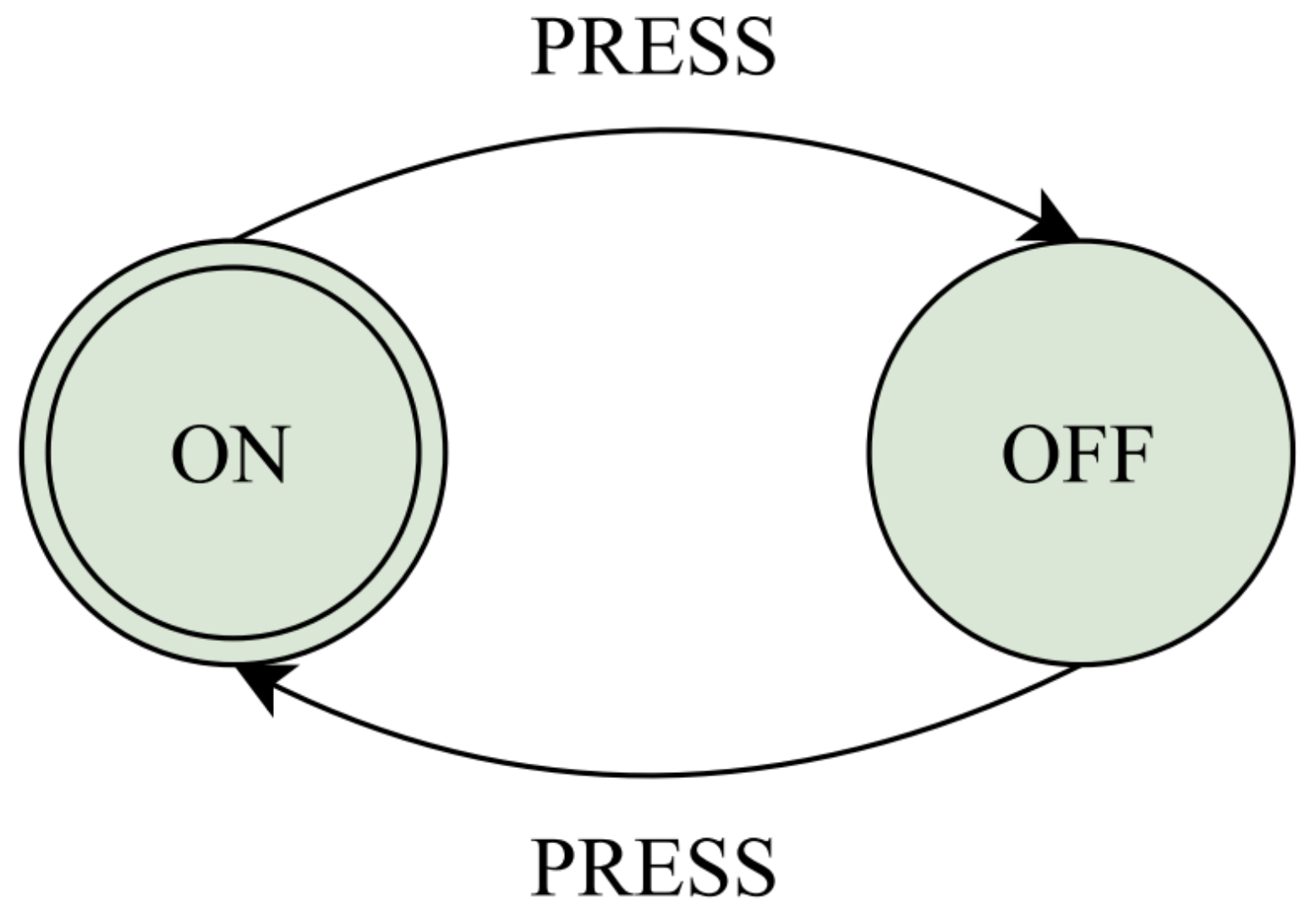}
    \end{minipage}


        \caption{
            Two-state finite-state machine modeling an on/off switch. The~initial state is ON (double ring). The~initial state is ON (double ring). Each PRESS triggers a transition to the other state.
        }
        \label{fig:FSM_toc_book}

\end{figure}

The machine only needs to remember ``whether it is in the on state or off state.'' From this, we can infer that a state represents the relevant part of a machine's history needed to predict a portion of the machine's future behavior. This raises the question: ``How to formalize this notion of state?'' Computational mechanics addresses this by formalizing the concept probabilistically, defining it as a triple, as~shown in Section~\ref{subsec:comp_mech}.

\section{Information Measures Across Varying Temperature in a Nearest-Neighbor Ising~Model} \label{appendix:nn_ising_info_vs_temp}

\begin{figure}[H]
    
    \begin{minipage}{0.55\textwidth}
        
        \includegraphics[width=\linewidth]{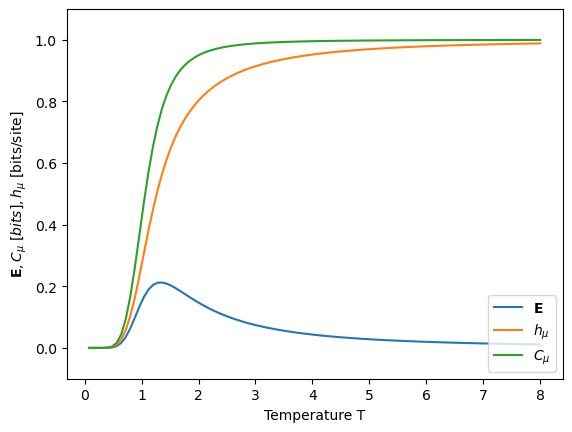}
    \end{minipage}

        \caption{$C_\mu$, $h_\mu$, and~$\mathbf{E}$ vs. $T$ for nn spin-$1/2$ Ising model with $B = 0.2$ and $J_1 = 1$. Adapted from Feldman and Crutchfield, 2022, \cite{feldman1998dnco}.}   
         \label{fig:nn_ising_model_info_measures_dnco}

\end{figure}

Figure~\ref{fig:nn_ising_model_info_measures_dnco} presents information measures $C_\mu$, $h_\mu$, and~$\mathbf{E}$ as functions of temperature $T$ for a nn Ising model with $B = 0.2$ and ferromagnetic coupling $J_1 = 1$. This figure reproduces Figure~13 of Ref.~\cite{feldman1998dnco}.

\unskip

\section{Shannon Entropy Density \boldmath{$h_{\mu}$} and Boltzmann Entropy Density $h_{\text{therm}}$} \label{appendix:hmu_versus_htherm}

\begin{figure}[H]

  \begin{minipage}{0.55\textwidth}
      \centering
      \includegraphics[width=\linewidth]{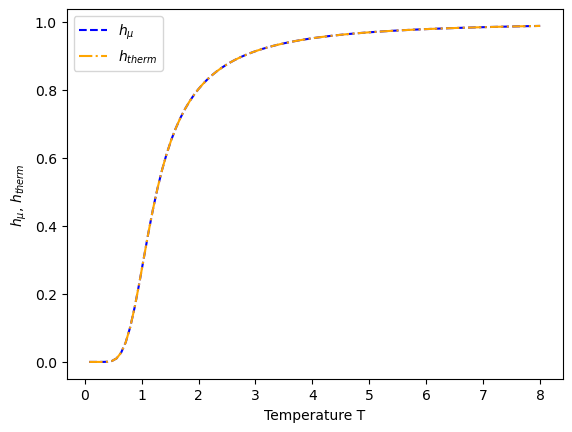}
  \end{minipage}


      \caption{Boltzmann (thermodynamic) entropy density $h_{\text{therm}}$ and Shannon entropy rate $h_{\mu}$ vs. temperature $T$.}  \label{fig:boltzmann_vs_shannon_entropy_rate}

\end{figure}

The form of the Boltzmann (thermodynamic) entropy density and Shannon entropy rate for a nn Ising model are presented in Equations~(\ref{eq:boltzmann_entropy_density}) and (\ref{eq:shannon_entropy_rate}), respectively. These expressions are plotted as a function of temperature $T$ in Figure~\ref{fig:boltzmann_vs_shannon_entropy_rate}, where they are graphically shown to be equivalent as temperature is varied.
\begin{equation}
h_{\text{therm}} = -\frac{\partial}{\partial T} \left( -\frac{T}{N} \log_2 \left( \lambda^N \right) \right)
\label{eq:boltzmann_entropy_density}
\end{equation}
\begin{equation}
h_\mu = - \sum_{s_0, s_{-1}=\pm1} \mathrm{Pr}(s_0, s_{-1}) \log_2 \left( \frac{\mathrm{Pr}(s_0, s_{-1})}{\mathrm{Pr}(s_{-1})} \right)
\label{eq:shannon_entropy_rate}
\end{equation}

\vspace{-15pt}
\unskip

\section{Derivation of Boltzmann (Thermodynamic) Entropy Density for Nearest-Neighbor Ising~Model} 
\label{appendix:nn_ising_thermo_entropy_derivation}

\begin{enumerate}
    \item Consider
    \begin{equation*}
    \begin{aligned}
    h_{\text{therm}} &= -\frac{\partial}{\partial T} \left( -\frac{T}{N} \log_2 \left( \lambda^N \right) \right) \\
    &=
    \frac{\partial}{\partial T} \left( T \log_2 \left( \lambda \right) \right)
    \\
    &= \log_2 \lambda + T \frac{\partial}{\partial T} (\log_2 \lambda)
    \end{aligned}
    \end{equation*}
    \item Using the chain rule:
    \begin{equation*}
    \begin{aligned}
    \frac{\partial}{\partial T} = \frac{d\beta}{dT} \frac{\partial}{\partial \beta} = -\frac{1}{T^2} \frac{\partial}{\partial \beta} = -\beta^2 \frac{\partial}{\partial \beta}
    \end{aligned}
    \end{equation*}
    \item Rewrite $h_{\text{therm}}$ in terms of $\beta=\frac{1}{T}
    $
    \begin{equation*}
    \begin{aligned}
    h_{\text{therm}} &= \log_2 \lambda - \beta \frac{\partial \log_2(\lambda)}{\partial \beta} \\
    &= \log_2 \lambda - \beta \frac{1}{\log(2)}\frac{1}{\lambda} \frac{\partial \lambda}{\partial \beta}
    \end{aligned}
    \end{equation*}
    \item Split principal eigenvalue $\lambda$ into two terms
    \begin{equation*}
    \begin{aligned}
    \lambda &= e^{\beta J} \cosh(\beta B) + \sqrt{e^{2\beta J} \sinh^2 (\beta B) + e^{-2\beta J}} \\
    &= \text{term I} + \text{term II} 
    \end{aligned}
    \end{equation*}
    \item Carry out $\frac{d}{d\beta} \text{term I}$ and $\frac{d}{d\beta} \text{term II}$ 
    \begin{equation*}
    \begin{aligned}
    \frac{d}{d\beta} \text{term I} &= e^{\beta J} \left( J \cosh(\beta B) + B \sinh(\beta B) \right) \\
    \frac{d}{d\beta} \text{term II} &= \frac{1}{2} \left( e^{2\beta J} \sinh^2 (\beta B) + e^{-2\beta J} \right)^{-\frac{1}{2}} \cdot \\
    & \frac{d}{d\beta} \left( e^{2\beta J} \sinh^2 (\beta B) + e^{-2\beta J} \right)
    \end{aligned}
    \end{equation*}
    \item Simplify $\frac{d}{d\beta} \left( e^{2\beta J} \sinh^2 (\beta B) + e^{-2\beta J} \right)$
    \begin{equation*}
    \begin{aligned}
    &= 2J e^{2\beta J} \sinh^2 (\beta B) + e^{2\beta J} \cdot
    2B \sinh (\beta B) \cosh (\beta B) - 2J e^{-2\beta J} \\
    &= 2 \left(J e^{2\beta J} \sinh^2 (\beta B) + B e^{2\beta J} \sinh (\beta B) \cosh (\beta B) - J e^{-2\beta J} \right) \\
    &= 2 \left(e^{-2\beta J} (J e^{4\beta J} \sinh^2 (\beta B) + B e^{4\beta J} \sinh (\beta B) \cosh (\beta B) - J) \right)
    \end{aligned}
    \end{equation*}
    \item Simplify $\frac{d}{d\beta} \text{term II}$
    \begin{equation*}
    \begin{aligned}
    &= \frac{J e^{2\beta J} \sinh^2 (\beta B) + B e^{2\beta J} \sinh (\beta B) \cosh (\beta B) - J e^{-2\beta J}}{\sqrt{e^{2\beta J} \sinh^2 (\beta B) + e^{-2\beta J}}}
    \end{aligned}
    \end{equation*}
    
    \begin{equation*}
    \begin{aligned}
    &= 2J e^{2\beta J} \sinh^2 (\beta B) + e^{2\beta J} \cdot
    2B \sinh (\beta B) \cosh (\beta B) - 2J e^{-2\beta J} \\
    &= 2 \left(J e^{2\beta J} \sinh^2 (\beta B) + B e^{2\beta J} \sinh (\beta B) \cosh (\beta B) - J e^{-2\beta J} \right) \\
    &= 2 \left(e^{-2\beta J} (J e^{4\beta J} \sinh^2 (\beta B) + B e^{4\beta J} \sinh (\beta B) \cosh (\beta B) - J) \right)
    \end{aligned}
    \end{equation*}
    \item Simplify $\frac{d \lambda}{d\beta}$
    \begin{equation*}
    \begin{aligned}
    \frac{d\lambda}{d\beta} &= e^{\beta J} \left( J \cosh(\beta B) + B \sinh(\beta B) \right) + \\
    & \frac{J e^{2\beta J} \sinh^2 (\beta B) + B e^{2\beta J} \sinh (\beta B) \cosh (\beta B) - J e^{-2\beta J}}{\sqrt{e^{2\beta J} \sinh^2 (\beta B) + e^{-2\beta J}}}
    \end{aligned}
    \end{equation*}
    \item Replace in $h_{\text{therm}}$
    \begin{equation*}
    \begin{aligned}
    h_{\text{therm}} &= \log_2 \lambda - \beta \frac{1}{\log(2)}\frac{1}{\lambda} \frac{\partial \lambda}{\partial \beta} \\
    &= \log_2 \lambda - \beta \frac{1}{\log(2)}\frac{1}{\lambda} \cdot ( e^{\beta J} \left( J \cosh(\beta B) + B \sinh(\beta B) \right) + \\
    & \frac{J e^{2\beta J} \sinh^2 (\beta B) + B e^{2\beta J} \sinh (\beta B) \cosh (\beta B) - J e^{-2\beta J}}{\sqrt{e^{2\beta J} \sinh^2 (\beta B) + e^{-2\beta J}}})
    \end{aligned}
    \end{equation*}

\end{enumerate}

\section{\boldmath{$\epsilon$}-Machine of Nearest-Neighbor Ising~Model} 
\label{appendix:nn_ising_machine}

\begin{figure}[H]
    
    \begin{minipage}{0.3\textwidth}

        \includegraphics[width=\linewidth]{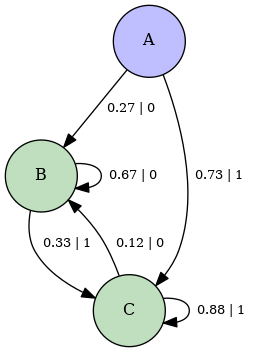}
    \end{minipage}
   
        \caption{
            $\epsilon$-machine 
 of nn Ising model with $J_1 = 1.0$, $B = 0.35$, and~$T = 1.5$.
        }
        \label{fig:nn_ising_model_machine_dnco}

\end{figure}

Figure~\ref{fig:nn_ising_model_machine_dnco} presents the $\epsilon$-machine of a nearest-neighbor Ising model with $J_1 = 1.0$, $B = 0.35$, and~$T = 1.5$. This figure reproduces Figure~10 of Ref.~\cite{feldman1998dnco}.

\unskip

\section{Joint Probability of Infinite~Chain} \label{appendix:joint_prob_of_infinite_chain}

\begin{enumerate}
    \item Consider a periodic infinite spin chain whose spins can only take two values (up or down) and only interact with their nearest neighbors.
    $$s_0, s_1, s_2, s_3, ..., s_{N-1} \ \text{where} \ s_0 = s_{N}$$
    \item Define a Hamiltonian for this system in a translation-invariant manner.
    $$E = E(s_0, \ldots, s_{N-1}) = -\sum\limits_{i=0}^{N-1} Js_is_{i+1} -\sum\limits_{i=0}^{N-1}  \frac{B(s_i+s_{i+1})}{2}$$
    \item Calculate the system's partition function. 
    $$Z= \sum\limits_{\{s_i\}}e^{-\beta E}$$
    \item Define the Boltzmann probability of a given infinite configuration.
    $$\mathrm{Pr}(s_0 \ldots s_{N}) = \dfrac{e^{-\beta E}}{Z}$$
    \item Define the transfer matrix matrix, with~components $V(s_i, s_{i+1}) = V_{s_is_{i+1}}=e^{-\beta E(s_i s_{i+1})}$.
    $$\mathbf{V} = \begin{bmatrix}
    e^{-\beta E(\uparrow, \uparrow)} & e^{-\beta E(\downarrow, \uparrow)} \\
    e^{-\beta E(\uparrow, \downarrow)} & e^{-\beta E(\downarrow, \downarrow)} \\
    \end{bmatrix}$$ 
    \item Express Boltzmann probability weight $e^{-\beta E}$ in terms of transfer matrix components.
    $$e^{-\beta E} = e^{-\beta E(s_0 s_1)}e^{-\beta E(s_1 s_2)} \ ... \ e^{-\beta E(s_{N-1} s_N)}$$
    $$= V_{s_0s_1} V_{s_1s_2} \ ... \ V_{s_{N-1}s_N}$$    
    \item Calculate partition function in the thermodynamic limit $N \rightarrow \infty$.
    $$Z_{N\rightarrow\infty} = \sum\limits_{s_0 = \pm 1} \sum\limits_{s_1 = \pm 1} \ ... \ \sum\limits_{s_N = \pm 1} V_{s_0s_1} V_{s_1s_2} \ ... \ V_{s_{N-1}s_N}$$
    \item Apply definition of matrix multiplication $\sum\limits_{s_2} V_{s_1 s_2} V_{s_2 s_3} = V^2_{s_1 s_3}$ and enforce periodic boundary conditions $s_0 = s_N$.
    $$Z_{N\rightarrow\infty} = \sum\limits_{s_0 = \pm 1} \sum\limits_{s_N = \pm 1} V^{N-1}_{s_0 s_N} = \sum\limits_{s_0 = \pm 1} V^N_{s_0 s_0}$$
    \item Apply definition of trace.
    $$Z_{N\rightarrow\infty} = (\mathrm{Tr}(\mathbf{V}))^N = \lim\limits_{N \rightarrow \infty} \lambda^N_{+} + \lambda^N_{-}$$
    $$= \lim\limits_{N \rightarrow \infty} \lambda^N_{+}(1 + \frac{\lambda^N_{+}}
    {\lambda^N_{-}})=\lambda^N_{+}=\lambda^N$$
    $$\text{where} \ \lambda \ \text{is principal eigenvalue}$$
    \item Express joint probability of a given infinite spin chain in terms of principal eigenvalue $\lambda$ and transfer matrix components $V_{s_i, s_{i+1}}$.
    $$\mathrm{Pr}(s_0, s_1, ..., s_{N-1}) = \dfrac{V_{s_0s_1} V_{s_1s_2} \ ... \ V_{s_{N-1}s_N}}{\lambda^N} = \dfrac{\prod\limits_{i=0}^{N-1} V_{s_is_{i+1}}}{\lambda^N}$$
\end{enumerate}

\section{Eigenvalue Decomposition of Transfer~Matrix} \label{appendix:eigevalue_decomposition_of_transfer_matrix}

\begin{enumerate}
    \item Express $\textbf{V}$ in terms of its eigenvalue decomposition
    $\mathbf{V} = \mathbf{UDU}^{-1}$.
    $$
    \textbf{V} = \begin{bmatrix}
    u_{+} & u_{-} \\
    u_{-} & -u_{+}
    \end{bmatrix} \begin{bmatrix}
    \lambda_{+} & 0 \\
    0 & \lambda_{-}
    \end{bmatrix} \begin{bmatrix}
    u_{+} & u_{-} \\
    u_{-} & -u_{+}
    \end{bmatrix}^{-1}
    $$
    $$
    = \lambda_{+} \begin{bmatrix}
    u_{+} & u_{-} \\
    u_{-} & -u_{+}
    \end{bmatrix} \begin{bmatrix}
    1 & 0 \\
    0 & \frac{\lambda_{-}}{\lambda_{+}}
    \end{bmatrix} \begin{bmatrix}
    u_{+} & u_{-} \\
    u_{-} & -u_{+}
    \end{bmatrix}^{-1}
    $$
    \item Use fact that in the thermodynamic limit $N \rightarrow \infty$, $\lambda_{+} \gg \lambda_{-}$. Rename $\lambda_{+}$ as $\lambda$.
    $$
    =  \begin{bmatrix}
    u_{+} & u_{-} \\
    u_{-} & -u_{+}
    \end{bmatrix} \begin{bmatrix}
    \lambda & 0 \\
    0 & 0
    \end{bmatrix} \begin{bmatrix}
    u_{+} & u_{-} \\
    u_{-} & -u_{+}
    \end{bmatrix}^{-1}
    $$
    $$
    = \begin{bmatrix}
    u_{+} & u_{-} \\
    u_{-} & -u_{+}
    \end{bmatrix} \begin{bmatrix}
    \lambda & 0 \\
    0 & 0
    \end{bmatrix} \begin{bmatrix}
    u_{+} & u_{-} \\
    u_{-} & -u_{+}
    \end{bmatrix}
    $$
    $$
    = \begin{bmatrix}
    u_{+} & u_{-} \\
    u_{-} & -u_{+}
    \end{bmatrix} \begin{bmatrix}
    \lambda u_{+} & \lambda u_{-} \\
    0 & 0
    \end{bmatrix}
    $$    
    Therefore,    
    $$
    \textbf{V} = \begin{bmatrix}
    \lambda u^2_{+} & \lambda u_{+}u_{-} \\
    \lambda u_{+}u_{-} & \lambda u^2_{-}  
    \end{bmatrix}
    $$
    \item Express transfer matrix components in terms of the principal eigenvalue $\lambda$ and the principal eigenvector components $u_{+}$ and $u_{-}$ at the thermodynamic limit.
\begin{equation} \label{eq:transfer_matrix_components}
    \begin{split}
    &V(\uparrow, \uparrow) = \lambda u^2_{+} \\
    &V(\uparrow, \downarrow) = V(\downarrow, \uparrow) = \lambda u_{+}u_{-} \\
    &V(\downarrow, \downarrow) = \lambda u^2_{-}
    \end{split}
    \end{equation}
\end{enumerate}

\section{Partition Function of Finite Chain with Fixed Boundary Conditions Embedded on Infinite~Chain} \label{appendix:fixed_bd_conds_partition_function}

\textbf{Base Case (\boldmath{$L=3$}):}
\begin{enumerate}
    \item Consider the partition function of a finite chain of length $3$ with fixed boundary conditions.
    $$Z_3 = \sum\limits_{s_1 = \pm 1} V(s_0^{\text{fix}}, s_1)V(s_1, s_2^{\text{fix}})$$
    $$= V(s_0^{\text{fix}}, \uparrow)V(\uparrow, s_2^{\text{fix}}) + V(s_0^{\text{fix}}, \downarrow)V(\downarrow, s_2^{\text{fix}})$$

    \item Express transfer matrix components in terms of principal eigenvalue and principal eigenvector components. For~simplicity, we will drop the $\mathcal{L}$ and $\mathcal{R}$, because~for the nn Ising model the left and right eigenvectors are the same.
    $$= \lambda u_{s_0}^{\text{fix}}u_{\uparrow} \cdot  \lambda u_{\uparrow}u_{s_2}^{\text{fix}} + \lambda u_{s_0}^{\text{fix}}u_{\downarrow} \cdot  \lambda u_{\downarrow}u_{s_2}^{\text{fix}}$$
    $$= \lambda u_{s_0}^{\text{fix}}u_{s_2}^{\text{fix}}u^2_{\uparrow} + \lambda u_{s_0}^{\text{fix}}u_{s_2}^{\text{fix}}u^2_{\downarrow}$$
    $$= \lambda u_{s_0}^{\text{fix}}u_{s_2}^{\text{fix}}(u^2_{\uparrow} + u^2_{\downarrow})$$
    $$= \lambda u_{s_0}^{\text{fix}}u_{s_2}^{\text{fix}}$$

\end{enumerate}

\textbf{Inductive Step:}
\begin{enumerate}
    \item Assume the partition function of a finite chain of length $L$ has the following expression.
\begin{equation}
        Z_L = \lambda^{L-1}u_{s_0}^{\text{fix}}u_{s_{L-1}}^{\text{fix}}
        \label{eq:part_func_L}
    \end{equation}
    \item Consider $Z_{L+1}$.
    $$Z_{L+1} = \sum\limits_{s_1 = \pm 1} \ldots \sum\limits_{s_{L-1} = \pm 1} V(s_0^{\text{fix}}, s_1) \ldots V(s_{L-1}, s_{L}^{\text{fix}})$$
    \item Sum over $L-1$.
\begin{align} \label{eq:part_func_L+1}
        =V(\uparrow, s_L^{\text{fix}}) \cdot \sum\limits_{s_1 = \pm 1} \ldots \sum\limits_{s_{L-2} = \pm 1} V(s_0^{\text{fix}}, s_1) \ldots V(s_{L-2}, \uparrow) \\ + V(\downarrow, s_L^{\text{fix}}) \cdot \sum\limits_{s_1 = \pm 1} \ldots \sum\limits_{s_{L-2} = \pm 1} V(s_0^{\text{fix}}, s_1) \ldots V(s_{L-2}, \downarrow)
        \notag
    \end{align}
    \item Replace Equation~(\ref{eq:part_func_L}) in Equation~(\ref{eq:part_func_L+1}).
\begin{equation}
        =V(\uparrow, s_L^{\text{fix}}) \cdot \lambda^{L-1} u_{s_0}^{\text{fix}}u_{\uparrow} + V(\downarrow, s_L^{\text{fix}}) \cdot \lambda^{L-1} u_{s_0}^{\text{fix}}u_{\downarrow}
        \label{eq:part_func_L+1_2}
    \end{equation}
    \item Replace Equation~(\ref{eq:transfer_matrix_components}) in Equation~(\ref{eq:part_func_L+1_2})
    $$=\lambda u_{\uparrow}  u_{s_L}^{\text{fix}} \cdot \lambda^{L-1} u_{s_0}^{\text{fix}}u_{\uparrow} + \lambda u_{\downarrow} s_L^{\text{fix}} \cdot \lambda^{L-1}u_{s_0}^{\text{fix}}u_{\downarrow}$$
    $$=\lambda^L u_{s_0}^{\text{fix}} u_{s_L}^{\text{fix}} u^2_{\uparrow} + u_{s_0}^{\text{fix}} u_{s_L}^{\text{fix}} u^2_{\downarrow} $$
    \item Factor.
    $$=\lambda^L u_{s_0}^{\text{fix}} u_{s_L}^{\text{fix}} (u^2_{\uparrow}+u^2_{\downarrow})$$
    \item Use normalization condition $u^2_{\uparrow}+u^2_{\downarrow} = 1$.
\begin{equation}
        Z_{L+1} = \lambda^L u_{s_0}^{\text{fix}} u_{s_L}^{\text{fix}}
    \end{equation}
\end{enumerate}

\section{Joint Probability of Finite Chain Embedded on Infinite~Chain} \label{appendix:joint_prob_of_embedded_finite_chain}
\begin{enumerate}
    \item Consider a finite spin chain embedded in an infinite spin chain.
    \begin{equation*}    \overrightarrow{s}^L = s_0, \ldots, s_{L-1}
   \end{equation*}
   
    \item The embedding of the finite spin chain~implies:
    \begin{itemize}
        \item The thermodynamic limit applies to the finite chain.
        \item 
        The magnetization is uniform across the bulk and boundaries of the finite chain.
    \end{itemize}
    
    \item To ensure uniform magnetization, express $\mathrm{Pr}_\text{embedded}$ in terms of conditional and marginal probabilities to separate the contributions from the bulk and boundaries. For~simplicity, we denote    $\mathrm{Pr}_\text{embedded}$ as $\mathrm{Pr}$.
    \begin{equation*}
    \mathrm{Pr}(\overrightarrow{s}^L)= \mathrm{Pr}(\overrightarrow{s}^L|s_0 \ \text{and} \ s_{L-1} \ \text{are fixed})\mathrm{Pr}(s_0, s_{L-1})  
    \end{equation*}

    \item Since $s_1$ and $s_L$ are independent, their probabilities can be factored as:
    \begin{equation*}        
    \mathrm{Pr}(\overrightarrow{s}^L)=\mathrm{Pr}(\overrightarrow{s}^L|s_0=s^{\text{fixed}}_0, s_{L-1}=s^{\text{fixed}}_{L-1})\mathrm{Pr}(s_0) \mathrm{Pr}(s_{L-1})   
    \end{equation*}
    \item Express $\mathrm{Pr}(\overrightarrow{s}^L|s_0 \ \text{and} \ s_L \ \text{are fixed})$ as a joint probability using $\mathrm{Pr}(s^{\text{fixed}}_i)=1$.   
    \begin{equation*}
        \mathrm{Pr}(\overrightarrow{s}^L|s_0 \ \text{and} \ s_L \ \text{are fixed}) = \mathrm{Pr}(s^{\text{fixed}}_0, \ldots, s^{\text{fixed}}_{L-1})
    \end{equation*}
    
    Thus,
\begin{equation}
        \mathrm{Pr}(\overrightarrow{s}^L)=\mathrm{Pr}(s^{\text{fixed}}_0, \ldots, s^{\text{fixed}}_{L-1})\mathrm{Pr}(s_0) \mathrm{Pr}(s_{L-1})
        \label{eq:embedded_prob_cond_form}
    \end{equation}
       
    \item Replace relevant joint and marginal probabilities for nn Ising model in Equation~(\ref{eq:embedded_prob_cond_form}).
    $$\mathrm{Pr}(\overrightarrow{s}^L) = \dfrac{\prod\limits_{i=0}^{L-2} V_{s_is_{i+1}}}{u_{\mathcal{L}, s_{0}} u_{\mathcal{R}, s_{L-1}}\lambda^{L-1}}\cdot u^2_{\mathcal{L}, s_{0}} \cdot u^2_{\mathcal{R}, s_{L-1}}$$
    $$ = \dfrac{u_{\mathcal{L}, s_{0}} u_{\mathcal{R}, s_{L-1}}\prod\limits_{i=0}^{L-2} V_{s_is_{i+1}}}{\lambda^{L-1}}$$ 
    \item To recover Equation~\eqref{eq:full_embedded_configuration}, consider $\overleftrightarrow{s}^L$ instead of $\overrightarrow{s}^L$

\end{enumerate}

\section{Finite-Range Ising Model Hamiltonian for \boldmath{$R=1, 2$} and $3$} \label{appendix:finite_range_ising_derivation}

The finite-range Ising model Hamiltonian is written below for neighborhood radii $R = 1$, $R = 2$, and~$R = 3$. In~the notation used for the Hamiltonian, the~neighborhood radius $R$ is denoted by $n$.

For $n = 1$:
\begin{equation*}
\begin{aligned}
X_{\eta_j} &= - B \sum\limits_{i=0}^{1-1} s_i^j - \sum\limits_{k=1}^{n=1} J_k \left( \sum\limits_{i=0}^{1-k-1} s_i^j s_{i+k}^j \right) \\
&= - B s_0^j - 0 = - B s_0^j \\
&= - B s_0
\end{aligned}
\end{equation*}

\begin{equation*}
\begin{aligned}
Y_{\eta_{j}, \eta_{j+1}} &= -\sum\limits_{k=1}^{n=1} J_k \left( \sum\limits_{i=0}^{k-1} s_{1-i-1}^j s_{k-i-1}^{j+1} \right) \\
&= -J_1 \left( \sum\limits_{i=0}^{1-1} s_{-i}^j s_{-i}^{j+1} \right) = -J_1 \left( s_0^j s_0^{j+1} \right) \\
&= -J_1 s_0 s_1 
\end{aligned}
\end{equation*}

\begin{equation*}
\begin{aligned}
X_{\eta_{j+1}} &= - B s_0^{j+1} = - B s_0^{j+1} = - B s_1
\end{aligned}
\end{equation*}

For $n = 2$:
\begin{equation*}
\begin{aligned}
X_{\eta_j} &= - B \sum\limits_{i=0}^{2-1} s_i^j - \sum\limits_{k=1}^{n=2} J_k \left( \sum\limits_{i=0}^{2-k-1} s_i^j s_{i+k}^j \right) \\
&= - B \left( s_0^j + s_1^j \right) - J_1 \left( \sum\limits_{i=0}^{2-1-1} s_i^j s_{i+1}^j \right) - J_2 \left( \sum\limits_{i=0}^{2-2-1} s_i^j s_{i+2}^j \right) \\
&= - B \left( s_0^j + s_1^j \right) - J_1 s_0^j s_1^j \\
&= - B \left( s_0 + s_1 \right) - J_1 s_0 s_1
\end{aligned}
\end{equation*}

\begin{equation*}
\begin{aligned}
Y_{\eta_{j}, \eta_{j+1}} &= -\sum\limits_{k=1}^{n=2} J_k \left( \sum\limits_{i=0}^{k-1} s_{2-i-1}^j s_{k-i-1}^{j+1} \right) \\
&= -J_1 \left( \sum\limits_{i=0}^{1-1} s_{1-i}^j s_{0-i}^{j+1} \right) - J_2 \left( \sum\limits_{i=0}^{2-1} s_{1-i}^j s_{1-i}^{j+1} \right) \\
&= -J_1 \left( s_1^j s_0^{j+1} \right) + J_2 \left( s_1^j s_1^{j+1} + s_0^j s_0^{j+1} \right) \\
&= -J_1 s_1 s_2 - J_2 \left( s_1 s_3 + s_0 s_2 \right)
\end{aligned}
\end{equation*}

\begin{equation*}
\begin{aligned}
X_{\eta_{j+1}} &= - B \left( s_0^{j+1} + s_1^{j+1} \right) - J_1 s_0^{j+1} s_1^{j+1} \\
&= - B \left( s_2 + s_3 \right) - J_1 s_2 s_3
\end{aligned}
\end{equation*}

For $n = 3$:
\begin{equation*}
\begin{aligned}
X_{\eta_j} &= - B \sum\limits_{i=0}^{3-1} s_i^j - \sum\limits_{k=1}^{n=3} J_k \left( \sum\limits_{i=0}^{3-k-1} s_i^j s_{i+k}^j \right) \\
&= - B \left( s_0^j + s_1^j + s_2^j \right) - J_1 \left( \sum\limits_{i=0}^{3-1-1} s_i^j s_{i+1}^j \right) \\
&- J_2 \left( \sum\limits_{i=0}^{3-2-1} s_i^j s_{i+2}^j \right) \\
&= - B \left( s_0^j + s_1^j + s_2^j \right) - J_1 \left( s_0^j s_1^j + s_1^j s_2^j \right) - J_2 s_0^j s_2^j \\
&= - B \sum\limits_{i=0}^{2} s_i - J_1 \left( s_0 s_1 + s_1 s_2 \right) - J_2 s_0 s_2
\end{aligned}
\end{equation*}

\begin{equation*}
\begin{aligned}
Y_{\eta_{j}, \eta_{j+1}} &= -\sum\limits_{k=1}^{3} J_k \left( \sum\limits_{i=0}^{k-1} s_{3-i-1}^{j} s_{k-i-1}^{j+1} \right) \\
&= -J_1 \left( \sum\limits_{i=0}^{1-1} s_{2-i}^{j} s_{0-i}^{j+1} \right) - J_2 \left( \sum\limits_{i=0}^{2-1} s_{2-i}^{j} s_{1-i}^{j+1} \right) \\
&- J_3 \left( \sum\limits_{i=0}^{3-1} s_{2-i}^{j} s_{2-i}^{j+1} \right) \\
&= -J_1 \left( s_2^j s_0^{j+1} \right) - J_2 \left( s_2^j s_1^{j+1} + s_1^j s_0^{j+1} \right) \\
& -J_3 \left( s_2^j s_2^{j+1} + s_1^j s_1^{j+1} + s_0^j s_0^{j+1} \right) \\
&= -J_1 \left( s_2 s_3 \right) - J_2 \left( s_2 s_4 + s_1 s_3 \right) \\
& -J_3 \left( s_2 s_5 + s_1 s_4 + s_0 s_3 \right)
\end{aligned}
\end{equation*}

\begin{equation*}
\begin{aligned}
X_{\eta_{j+1}} &= - B \left( s_0^{j+1} + s_1^{j+1} + s_2^{j+1} \right) \\
&- J_1 \left( s_0^{j+1} s_1^{j+1} + s_1^{j+1} s_2^{j+1} \right) \\
&- J_2 s_0^{j+1} s_2^{j+1} \\
&= -B \sum\limits_{i=3}^{n=5} s_i - J_1 \left( s_3 s_4 + s_4 s_5 \right) - J_2 s_3 s_5
\end{aligned}
\end{equation*}

\section{Three-Body Model Transfer~Matrix} \label{appendix:three_body_transfer_matrix}

The transfer matrix $\mathbf{V}$ of the three-body spin model is shown in Equation~\eqref{eq:3body_transfer_matrix}. To~simplify the notation of the matrix entries, we label each two-spin block as follows: $\uparrow\uparrow \ = 1$, $\uparrow\downarrow \ = 2$, $\downarrow\uparrow \ = 3$, and~$\downarrow\downarrow \ = 4$. Moreover, we set the chemical potential $\mu$ to zero.
\begin{equation}
\mathbf{V} = 
\begin{array}{c@{\hspace{0.5em}}c}
    & \begin{array}{@{\hspace{0.6em}}c@{\hspace{1em}}c@{\hspace{1em}}c@{\hspace{1em}}c@{\hspace{0.6em}}} \uparrow\uparrow & \uparrow\downarrow & \downarrow\uparrow & \downarrow\downarrow \end{array} \\
    \begin{array}{c} \uparrow\uparrow \\ \uparrow\downarrow \\ \downarrow\uparrow \\ \downarrow\downarrow \end{array} &
    \begin{bmatrix}
        V_{11} & V_{12} & 0 & 0 \\
        0 & 0 & V_{23} & V_{24} \\
        V_{31} & V_{32} & 0 & 0 \\
        0 & 0 & V_{43} & V_{44} \\
    \end{bmatrix}
\end{array}
\label{eq:3body_transfer_matrix}
\end{equation}
where
\begin{equation*} \\
\begin{aligned}
V_{11} &= \exp\left(\frac{\mu - J_1 - J_2 - J_\text{tb}}{T}\right), \\
V_{12} &= \exp\left(\frac{\mu - J_1}{T}\right), \\
V_{23} &= \exp\left(\frac{\mu - J_2}{T}\right), \\
V_{24} &= \exp\left(\frac{\mu}{T}\right), \\
V_{31} &= V_{32} = V_{43} = V_{44} = 1
\end{aligned}
\end{equation*}

Note that, unlike the finite-range Ising model, the~spin blocks in this model overlap by one spin. Specifically, the~last spin in a row label must match the first spin in a column label. For~example, the~spin block $\uparrow \downarrow$ in the second row can only transition to spin blocks $\downarrow \uparrow$ or $\downarrow \downarrow$ in the third and fourth columns, as~its last spin $\downarrow$ matches the first spin of both $\downarrow \uparrow$ and $\downarrow \downarrow$.

\begin{adjustwidth}{-\extralength}{0cm}
\reftitle{References}

\PublishersNote{}
\end{adjustwidth}

\end{document}